\documentclass{aa}
\usepackage{graphicx}
\usepackage{txfonts}
\usepackage{natbib}
\usepackage{color}
\bibpunct{(}{)}{;}{a}{}{,}    
\newcommand{\beq}{\begin{equation}}
\newcommand{\eeq}{\end{equation}}
\newcommand{\bea}{\begin{eqnarray}}
\newcommand{\eea}{\end{eqnarray}}

\newcommand{\url}[1]{{\tt #1}}

\newcommand{\doverd}[2]{\frac{\partial #1}{\partial #2}}

\def\gapp{\lower 3pt\hbox{${\buildrel > \over \sim}$}\ }
\def\lapp{\lower 3pt\hbox{${\buildrel < \over \sim}$}\ }

\usepackage{color}
\usepackage{ulem}

\newlength{\linwx}
\begin{document}
\title{Stellar irradiated discs and implications on migration of embedded planets I: equilibrium discs}
\author{
Bertram Bitsch \inst{1},
Aur\'{e}lien Crida \inst{1},
Alessandro Morbidelli \inst{1},
Wilhelm Kley  \inst{2}
\and
Ian Dobbs-Dixon \inst{3}
}
\offprints{B. Bitsch,\\ \email{bertram.bitsch@oca.eu}}
\institute{
University of Nice-Sophia antipolis, CNRS, Observatoire de la C\^{o}te d'Azur,
Laboratoire LAGRANGE, BP4229, 06304 NICE cedex 4, FRANCE
\and
     Institut f\"ur Astronomie \& Astrophysik, 
     Universit\"at T\"ubingen,
     Auf der Morgenstelle 10, D-72076 T\"ubingen, Germany
\and
Department of Astronomy, Box 351580, University of Washington, Seattle, WA 98195
}
\abstract
{ The strength and direction of migration of embedded low mass planets depends on the disc's thermodynamic state. It has been shown that, in discs where the viscous heating is balanced by radiative transport, the migration can be directed outwards, a process which extends the lifetime of growing planetary embryos.
 } 
{ In this paper we investigate the influence of opacity and stellar irradiation on the disc thermodynamics. We focus on equilibrium discs, which have no net mass flux. Utilizing the resulting disc structure, we determine the regions of outward migration in the disc.
} 
{ We perform two-dimensional numerical simulations of equilibrium discs with viscous heating, radiative cooling and stellar irradiation. We use the explicit/implicit hydrodynamical code {{\tt NIRVANA}} that includes a full tensor viscosity and stellar irradiation, as well as a two temperature solver that includes radiation transport in the flux-limited diffusion approximation. The migration of embedded planets is studied by using torque formulae.
 } 
{
In the constant opacity case, our code reproduces the analytical results corresponding to a black-body disc: the stellar irradiation dominates in the outer regions -- leading to flaring ($H/r \propto r^{2/7}$) -- while the viscous heating dominates close to the star. In particular, we find that the inner edge of the disc should not be significantly puffed-up by the stellar irradiation. If the opacity depends on the local density and temperature, the structure of the disc is different, and several bumps in the aspect ratio $H/r$ appear, due to transitions between different opacity regimes. The bumps in the disc structure are very important, as they can shield the outer disc from stellar irradiation.
}
{
 Stellar irradiation is an important factor for determining the disc structure and has dramatic consequences for the migration of embedded planets. Compared to discs with only viscous heating and radiative cooling, a stellar irradiated disc features a much smaller region of outward migration for a range of planetary masses. This suggests that the region where the formation of giant planet cores takes place is smaller, which in turn might lead to a shorter growth phase.
}
\maketitle
\markboth
{Bitsch et al.: Stellar irradiated discs and implications on migration of embedded planets}
{Bitsch et al.: Stellar irradiated discs and implications on migration of embedded planets}

\section{Introduction}
\label{sec:introduction}
In accretion discs around young stars planetary embryos are born and grow through collisions to form bigger and bigger objects. The largest protoplanets constitute the cores of giant planets. Embedded planets interact with the gas in the disc and can move through the disc. This process depends on the disc's physical properties \citep{1997Icar..126..261W}, in particular the local density and temperature and the gradients thereof.

In a real disc, not only viscous heating and radiative cooling determine the disc's thermal structure, but also stellar irradiation. In the outer parts of the disc the heating from the star can keep the disc flared \citep{1997ApJ...490..368C}, in contrast to the disc profile without stellar irradiation. The effect of stellar irradiation on the disc structure has largely been investigated utilizing a 1D+1D numerical approach (e.g. \citet{1997ApJ...486..372B,2001ApJ...560..957D}) with the goal of fitting the spectral energy distributions (SEDs) of observed discs

The migration of a low-mass planet embedded in a fully radiative gaseous disc can be significantly different from migration in an isothermal or purely adiabatic disc \citep{2006A&A...459L..17P, 2008ApJ...672.1054B, 2008A&A...485..877P, 2008A&A...478..245P, 2008A&A...487L...9K, 2009A&A...506..971K, 2010MNRAS.408..876A}. While all authors agree that radiation transport can slow the rate of inward migration, there is still a lack of consensus whether the direction of migration can be outward and, if so, in which part of the disc \citep{2011A&A...536A..77B}. Part of this confusion may be due to the sensitivity of the direction and magnitude of migration on local disc parameters \citep{2010MNRAS.401.1950P, 2010ApJ...723.1393M, 2011MNRAS.410..293P}, including, for example, the radial disc temperature gradient \citep{2011MNRAS.415..576A}.

In all previous works on planetary migration, the disc structure is determined by viscous heating and cooling. The equilibrium disc structure is thus determined by the disc mass and the magnitude of viscosity \citep{2011A&A...536A..77B}. Including stellar irradiation can lead to a flared disc profile in the outer parts of the disc \citep{2004A&A...417..159D}. The resulting change in the disc structure in the outer parts can have a huge effect on the migration of embedded planets. As the aspect ratio of the disc $H/r$ changes, so does the temperature profile, which in turn determines the strength of the entropy gradient. The entropy gradient determines the magnitude of the corotation torque \citep{2008ApJ...672.1054B}, which constitutes a significant fraction of the total torque, and hence determines the strength and direction of migration.

In this paper we investigate the structure of the disc due to changes in opacity and the effects of stellar irradiation using a 2D ($r-z$) disc model. In Section \ref{sec:methods} we describe the energy equation including stellar heating and the opacities used. We then compare test studies with analytical calculations of the disc structure in the case of constant opacity in Section \ref{sec:constopc}, followed by studies with non constant opacity in Section \ref{sec:nconstopc}. The differences between discs including stellar heating and those with only viscous heating are described in Section \ref{sec:discs}. We then apply torque formulae to estimate the torque acting on embedded planets for these disc structures in Section \ref{sec:migration}. In Section \ref{sec:summary} the summary and conclusions are presented.

\section{Methods}
\label{sec:methods}
 
The protoplanetary disc is treated in this study as a two-dimensional (2D) non-self-gravitating gas whose motion is described by the Navier-Stokes equations. In this paper we focus on the axisymmetric structure of the disc and compute a vertical slice (in the $r-z$ plane) utilizing 2D spherical coordinates ($r$-$\theta$). Turbulence in discs is thought to be driven by magneto-hydrodynamical instabilities \citep{1998RvMP...70....1B} but here we treat viscosity utilizing either a constant kinematic viscosity or an $\alpha$-viscosity \citep{1973A&A....24..337S}. The dissipative effects can then be described via the standard viscous stress-tensor approach \citep[eg.][]{1984frh..book.....M}. We also include the irradiation from the central star, which will be described in detail in section.~\ref{subsec:Energy}. 
For that purpose we modified and substantially extended an existing multi-dimensional hydrodynamical code {\tt Nirvana} \citep{1997ZiegYork,2001ApJ...547..457K}. 

The radiative energy associated with viscous heating and stellar irradiation is then diffused through the disc and emitted from its surfaces. To describe this process we utilize the flux-limited diffusion approximation \citep[FLD,][]{1981ApJ...248..321L}, an approximation that allows us to transition from the optically thick mid-plane to the thin regions near the disc's surface.

The hydrodynamical equations solved in the code have already been described in detail \citep{2009A&A...506..971K}, so we focus here on our newly implemented two-temperature approach in Section~\ref{subsec:Energy} and how to solve the coupled equations in Appendix~\ref{ap:energy}.

\subsection{Energy equation}
\label{subsec:Energy}

In the two-temperature approach (featuring the radiative energy density $E_R$ and the thermal energy density $\epsilon$) the evolution equations for the thermal and radiation energy read \citep{1989A&A...208...98K, 2010ApJ...710.1395D, 2011A&A...529A..35C}:  
\begin{eqnarray}
\label{eq:energy}
 \doverd{\, E_R}{t} + \nabla \cdot \mathbf{F} &=& \rho \kappa_P (T,P) [B(T) - c E_R]  \\
 \doverd{\, \epsilon}{t} + (\mathbf{u} \cdot \nabla) \epsilon &=&  - P \nabla \cdot \mathbf{u} - \rho \kappa_P (T,P) [B(T) - c E_R] + Q^+ + S \nonumber.
\end{eqnarray}
Here, $E_R$ is independently evolved from the thermal component with $B(T) = 4 \sigma T^4$ ($\sigma$ being the Stefan–Boltzmann constant and $T$ the temperature of the gas). $\mathbf{F}$ denotes the radiative flux. From eq.~\ref{eq:energy}, one can calculate the coupling time-scale $\tau_{coup}=1/(\rho \kappa c)$ with the density $\rho$, the opacity $\kappa$, and $c$ the speed of light. The coupling time-scale defines the coupling between $B(T)/c$ and $E_R$. If $\rho \kappa$ is high, the coupling time is small, indicating that $E_R$ relaxes quickly towards $B(T)/c$, meaning $E_R \approx 4 \sigma T^4 /c$ (optically thick limit). If $\rho \kappa$ is low, the coupling time is large, and $E_R$ and $B(T)/c$ can decouple (optically thin limit). $B(T) - cE_R$ represents the exchange of energy between the thermal and radiative components through emission and absorption of low energy photons. $\kappa_P$ is the Planck opacity (see section~\ref{subsec:opac}). The thermal energy $\epsilon$ is given by $\epsilon=\rho c_v T$ ($c_v$ being the specific heat at constant volume which we hold fixed and uniform), and $\mathbf{u} = (u_r, u_\theta, u_\varphi)$ the gas velocity. $P$ is the gas pressure, $Q^+$ the viscous dissipation function and $S$ the contribution from the stellar heating.

The radiative flux, using flux-limited diffusion (FLD), can be written as
\begin {equation}
\label{eq:raddif}
 \qquad   \vec{F}  =  -  \frac{\lambda c}{\rho \kappa_R} \, \nabla E_R\ ,
\end{equation}
where $\kappa_R$ is the Rosseland mean opacity and $\lambda$ the flux limiter. Please note that $\kappa_P$ and $\kappa_R$ are different opacities. More details concerning the opacities used here are given in Section \ref{subsec:opac}. Using FLD allows us to solve for stable accretion disc models that cover several vertical pressure scale heights. We use here the FLD approach described in \citet{1981ApJ...248..321L} with the flux-limiter of \citet{1989A&A...208...98K} given by
\begin{equation}
\lambda =  \left\{
    \begin{array}{cc} 
    \frac{2}{3+\sqrt{9+10 R^2}}
     \quad &  \mbox{for} \quad  R \leq 2.0  \\
    \frac{10}{10R + 9 + \sqrt{81+180 R}} \quad & \mbox{for} \quad R > 2.0 
    \end{array}
    \right. ,
\end{equation}
where
\begin{equation}
 R = \frac{1}{\rho \kappa_R} \frac{|\nabla E_R|}{E_R} \ .
\end{equation}
These are the same specifications as in \citet{2009A&A...506..971K}, who used a one-temperature approach with $E_r = a_r T^4$, where $a_r = 4 \sigma/c$ is the radiation constant.

For our purposes the physical extent of the star is small with respect to the disc extension so we approximate it as a point source. In spherical coordinates, the stellar radiation is thus propagated along the radial direction only. With this method, the code simultaneously treats the stellar heating component via ray-tracing and subsequent re-radiation via FLD. The stellar heating density $S$ (i.e. the energy deposited per unit time and volume), received by a grid cell $i$ of width $\Delta r$ is given by:
\begin{equation}
 S = F_\star  e^{-\tau_i} \frac{1-e^{-\rho \kappa_\star \Delta r}}{\Delta r} \quad \mbox{with} \quad F_\star = \frac{R_\star^2 \sigma T_\star^4}{r^2}  \ ,
\end{equation}
where $R_\star$ is the stellar radius and $T_\star$ the stellar surface temperature. The factor $e^{-\tau_i}$ expresses how much stellar irradiation has been absorbed in the grid before it hits the radial grid cell $i$, with $\tau_i$ being the optical depth integrated radially towards grid cell $i$. The factor $1-e^{-\rho \kappa_\star \Delta r}$ describes how much stellar irradiation is absorbed in the actual grid cell, with $\kappa_\star$ being the optical opacity, calculated using the stellar spectrum \citep{2010ApJ...710.1395D}.
For more information regarding the opacities, please see Section.~\ref{subsec:opac}. In the optically thin limit ($\rho \kappa_\star \Delta r < 1.0$), we make the following approximation:
\begin{equation}
\label{eq:stellarthin}
 S = F_\star e^{-\tau_i} \rho \kappa_\star \ .
\end{equation}
More details concerning the stellar heating function are given in Appendix~\ref{ap:heat}.

The coupled energy equations (eq.~\ref{eq:energy}) have to be solved simultaneously and we follow here the approach outlined in \citet{2011A&A...529A..35C}. The advection $(\mathbf{u} \cdot \nabla) \epsilon$ and the compressional heating $(-P\nabla \cdot \mathbf{u})$ terms are solved in separate routines. Appendix~\ref{ap:energy} describes our approach to simultaneously solving the coupled energy equations (eq.~\ref{eq:energy}).

\subsection{Physical setup}
\label{subsec:phys}

\subsubsection{Computational Parameters}

For computational reasons, the inner disc ($0.0416$AU $< r < 1.04$AU) and outer disc ($1.04$AU $< r < 50.96$AU) are treated in two different sets of simulations. Both sets of simulations are 2D discs in the $r$-$\theta$ direction (radial and vertical) with $384 \times 32$ (inner disc) or $384 \times 64$ (outer disc) active grid cells. The opening angle used by the grid, i.e. the range of $\theta$, differs for the inner and outer disc simulations. In the inner disc $83^\circ < \theta < 90^\circ$, while in the outer disc $70^\circ < \theta < 90^\circ$, where $\theta$ is the colatitude measured from the $z$-axis and $\theta=90^\circ$ is the mid-plane.

The reason for introducing two sets of simulations for the inner and outer disc is numerical. The time step for the inner disc, due to the small inner radius, is very small, significantly increasing the computation time. It is therefore not feasible to simulate the whole disc in one simulation. Additionally, the opening angle of the numerical disc needs to increase to larger values at larger radial distances as the initial disc structure is flared ($H/r$ increasing with $r$). Setting an opening angle too large (e.g. $20^\circ$) for the inner disc results in a collapse of the time step due to very low densities in the upper regions of the disc. Test runs using a density floor showed the same behaviour. Therefore, the two sets of simulations utilize different opening angles. The simulations then can be attached to each other at the respective boundaries of the numerical grids, resulting in a continuous profile. The transfer of the stellar heating is described more precisely in section \ref{subsubsec:stellheat}.

We assume the upper and lower parts of the disc are identical and therefore apply symmetric boundary conditions at the disc's mid-plane, i.e. at $\theta_{max}=90^\circ$. At the top of the disc (at $\theta_{min}$) we set the radiation energy to $E_R = a_R T^4$ with $T=3.0$K (the temperature of the interstellar medium). In this way the disc will always be cooled by the upper boundary and all the heating created in the disc (viscous and stellar) can be transported outwards, as the boundary radiation energy is always lower than in mid-plane. This type of cooling at the top of the disc has been used in \citet{2009A&A...506..971K}, and is also adopted in SPH simulations of fully radiative discs \citep{2011MNRAS.415..576A}. In the radial direction we use reflecting boundary conditions for all disc parameters as described in \citet{2009A&A...506..971K}.

We start the simulations with an initial surface density profile of $\Sigma = \Sigma_0 (r/1$AU$)^{-1/2}$, where $\Sigma_0 = 1000$g/cm$^2$ or $\Sigma_0 = 3000$g/cm$^2$, which is approximately the value used in \citet{1977Ap&SS..51..153W} at $1$AU. The initial aspect ratio $H/r \propto r^{2/7}$ indicates a flared disc, following the ''flaring disc principle'' \citep{2002A&A...395..853D}: if the disc can flare, it will. Self-shadowed discs are discs that cannot be made to flare, which may occur for discs that are initialized with constant $H/r$.

We use either a constant kinematic viscosity of $\nu = 10^{15}$\,cm$^2$/s or an $\alpha$ prescription with different values for $\alpha$, following \citet{1973A&A....24..337S} with $\nu = \alpha c_s^2 / \Omega_K$, where $c_s$ is the mid-plane sound speed and $\Omega_K$ the Keplerian orbital frequency. In case of $\alpha$-viscosity we adopt a vertically constant viscosity utilizing the mid-plane values. MHD simulations indicate that the turbulent stresses do not strongly change with height \citep{2012MNRAS.420.2419F}. For discs with constant viscosity, the initial surface density profile is the equilibrium profile, which cancels viscous transport. For $\alpha$-discs the profile will evolve until a new equilibrium profile is achieved under the influence of reflecting boundary conditions. In this way, the mass inside the disc is conserved. This case is what we call equilibrium disc.

\subsubsection{Stellar heating}
\label{subsubsec:stellheat}

The star at the centre of our grid has $M_\star = M_\odot$, $T_\star=5656$K and $R_\star = 3.0 R_\odot$, corresponding to a typical protostar. In our simulations the inner discs starts at $r = 0.0416$AU, close to the corotation radius of the star. 

Let us first assume that the inner edge of the disc is sharply cut off and it cools through that edge as a blackbody. If all of the stellar irradiation is absorbed at this edge, the received energy per time is 
\begin{equation}
 \tilde{S}  = 2 \pi r_{min} 2 H F_\star = 4 \pi r_{min} H \frac{\sigma T_\star^4 R_\star^2}{r_{min}^2} = 4 \pi \sigma T_\star^4 R_\star^2 \frac{H}{r_{min}} \ ,
\end{equation}
where $2 \pi r_{min} 2 H$ is the surface of the inner face of the disc. $r_{min}$ denotes the radius of the inner edge of the disc. The blackbody cooling of the surface of the inner edge is given by
\begin{equation}
 Q_- = - 4 \pi r_{min} H \sigma T_D^4 \ ,
\end{equation}
with $T_D$ being the disc's temperature. In equilibrium (heating is equal to cooling) we get
\begin{eqnarray}
 \tilde{S} + Q_- = 0 \quad &\Rightarrow \quad 4 \pi \sigma T_\star^4 R_\star^2 \frac{H}{r_{min}} - 4 \pi r_{min} H \sigma T_D^4 = 0 \nonumber \\
 &\Rightarrow \quad T_\star^4 R_\star^2 = T_D^4 r_{min}^2.
\end{eqnarray}
Using vertical hydrostatic equilibrium,
\begin{equation}
 T_D = \left( \frac{H}{r_{min}} \right)^2 \frac{G M_\star}{r_{min}} \frac{\mu}{{\cal R}} \ ,
\end{equation}
with $\mu$ being the mean molecular weight and $\cal R$ the gas constant, we obtain
\begin{eqnarray}
\label{eq:Hrinner}
 T_\star^4 R_\star^2 &=& \left(\frac{H}{r_{min}}\right)^8 \frac{G^4 M_\star^4}{r_{min}^4} \frac{\mu^4}{{\cal R}^4} r_{min}^2 \nonumber \\
 \left( \frac{H}{r_{min}} \right)^8 &=& \frac{T_\star^4 R_\star^2 {\cal R}^4}{G^4 M_\star^4 \mu^4} r_{min}^2 = C_\star r_{min}^2 \nonumber \\
 \Rightarrow \quad \frac{H}{r_{min}} &=& C_\star^{1/8} r_{min}^{1/4} \\ 
  &=& 0.0516 \left( \frac{M_\star}{M_\odot} \right)^{-1/4} \left( \frac{R_\star}{3 R_\odot} \right)^{1/4} \left( \frac{T_\star}{5600 {\rm K}} \right)^{1/2}    \left( \frac{r_{min}}{1 {\rm AU}} \right)^{1/4} \nonumber \ .
\end{eqnarray}
Eq.~\ref{eq:Hrinner} indicates that the aspect ratio of the inner rim scales with the grid truncation radius ($r_{min}$) as $r_{min}^{0.25}$; decreasing $r_{min}$ results in a decreased $H/r$. Therefore, regardless of our choice of $r_{min}$, provided that the chosen value of $r_{min}$ is smaller than the radius of the innermost bump in $H/r$ due to opacity transitions (see Section~\ref{sec:nconstopc}), we are guaranteed that the disc \textit{interior} to this point will not have a larger aspect ratio than that at the chosen truncation radius. This has important consequences for the irradiation received by this first cell, as it guarantees this first cell will not be in the shadow of a higher $H/r$ region interior to it. Though the simulation naturally captures self-shadowing in the simulated regions of the disc, eq.~\ref{eq:Hrinner} shows this is not relevant for the region interior to $r_{min}$. Therefore, if $r_{min}$ is larger than the physical truncation radius of the disc, we can mimic the absorption of stellar irradiation due to the inner disc by simply reducing the stellar irradiation by a factor of $e^{-\tau_{inner}}$ before it hits the first active grid cell. In our case $\tau_{inner}$ corresponds to the optical depth of the inner two ghost cells of the numerical grid, that are located inside of $r_{min}$. These ghost cells have the same properties (e.g. $\rho$, $\kappa$, $\Delta r$) as the first active cell. By choosing the values of the ghost cells for $e^{-\tau_{inner}}$, we follow the radial profile of the disc, making $\tau_{inner}$ consistent with the rest of the simulation.

As some of the stellar irradiation is absorbed by the inner disc, the outer disc will not receive stellar irradiation in the optically thick mid-plane region. From the simulations of the inner disc, we can measure to what angle from mid-plane in the disc the stellar irradiation is absorbed, $\vartheta_{abs}$, by looking at the distribution of stellar heating in the inner disc. This angle $\vartheta_{abs}$ varies with the disc mass and viscosity. In the outer disc simulation, the disc will only receive stellar heating for an opening angle $\vartheta_o$ larger than $\vartheta_{abs}$ (keep in mind that $\vartheta=\theta - 90^\circ = 0$ represent the mid-plane of the disc):
\begin{equation}
\label{eq:thetaabs}
F_\star =  \left\{
    \begin{array}{cc} 
    F_\star \quad &  \mbox{for} \quad  \vartheta_o \geq \vartheta_{abs}  \\
    0.0 \quad & \mbox{for} \quad \vartheta_o < \vartheta_{abs} 
    \end{array}
    \right. .
\end{equation}
For the simulations of the outer disc, we reduce the stellar irradiation on the top layers by $e^{-\tau_{inner}}$ as we did for the simulations of the inner disc. This has the effect that the transition from zero stellar irradiation to stellar irradiation is smoothed out a little bit. This, of course, implies that the simulations of the inner disc have to be done before the simulations of the outer disc to obtain the correct $\vartheta_{abs}$.

In principle one could radially integrate the stellar flux from the inner boundary of the inner simulation to the outer boundary to see what remains. This would then act as the starting stellar heating of the simulations of the outer grid. However, as the simulations of inner and outer disc have different opening angles, this could only be done up to the opening angle of the inner disc. Also, as the inner parts are (radially) optically thick, most of the stellar irradiation is absorbed in the first cells, resulting in a zero stellar irradiation at the outer boundary of the inner disc simulations. Only the very top layers of the inner disc simulations still receive stellar irradiation at their outer boundary, which can be mimicked by the introduction of $\vartheta_{abs}$. In Appendix~\ref{ap:match} we present test simulations showing that the inner and outer disc simulations match perfectly with our simple prescription of $\vartheta_{abs}$.

\subsection{Opacities}
\label{subsec:opac}

In the previous sections we have introduced three different opacities, the Planck mean opacity $\kappa_P$, the optical opacity $\kappa_\star$ and the Rosseland mean opacity $\kappa_R$. In principle the opacities can be calculated as in \citet{2010ApJ...710.1395D}. The Rosseland and Planck mean opacities are defined in the usual manner, so the local Planck mean opacity for the low-energy photon group is given by
\begin{equation}
 \kappa_P (T, \rho) = \frac{\int \kappa_{\nu,ns}(T,\rho) B_\nu (T) d\nu}{\int B_\nu (T) d\nu} \ ,
\end{equation}
while the optical opacity is given by the high-energy photon group
\begin{equation}
 \kappa_\star (T, \rho) = \frac{\int \kappa_{\nu,ns}(T,\rho) J_\nu (T_\star) d\nu}{\int J_\nu (T_\star) d\nu} \ .
\end{equation}
The frequency dependent opacity is given by $\kappa_\nu$ and the subscript $ns$ indicates scattering processes, which are neglected when calculating wavelength dependent opacities. In principle, the spectrum of impinging radiation can differ from a blackbody, but for our purposes we set $J_\nu (T_\star) = B_\nu (T_\star)$. Finally, the Rosseland mean for the low-energy group is given by
\begin{equation}
 \kappa_R (T, \rho)^{-1} = \frac{\int \kappa_{\nu,s}^{-1} (T,\rho) \frac{\partial B_\nu (T)}{dT} d\nu}{\int \frac{\partial{B_\nu (T)}}{dT} d\nu} \ .
\end{equation}
The wavelength-dependent opacities include the effect of scattering (subscript $s$). \citet{2010ApJ...710.1395D} state that the ratio between $\kappa_\star$ and $\kappa_P$ is about a factor of $10$. They calculated the opacities directly from wavelength-dependent opacities using gaseous, solar-composition opacities. The net result of their work is that the photospheres for the stellar irradiation and the cooling are physically disconnected. The upper layers of the disc (where stellar irradiation plays the largest role) is largely depleted of dust, particularly at a later stage of evolution, when planets are forming. Thus the opacity due to grains is reduced (as is done in \citet{2009MNRAS.393...49A}). In our first set of simulations, we leave the Rosseland and Planck opacities constant at $1.0$cm$^2$/g, but the optical opacity is reduced relative to the other used opacities by a factor of $10$, so that $\kappa_\star=0.1$cm$^2$/g. 

\begin{figure}
 \centering
 \resizebox{\hsize}{!}{\includegraphics[width=0.9\linwx]{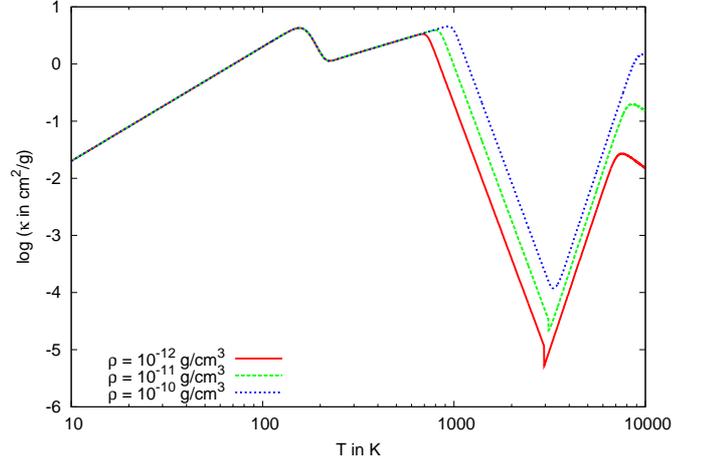}}
 \caption{Opacity from \citet{1994ApJ...427..987B} on a logarithmic scale. Different colours indicate different densities. The bumps in the profile originate from transitions in the opacity regime. This plot features $\kappa_R$ and $\kappa_P$ as we assume in this paper that $\kappa_R = \kappa_P$.
   \label{fig:opacity}
   }
\end{figure}

For simulations with varying opacities, we follow the opacity law of \citet{1994ApJ...427..987B}. The opacity depends primarily on temperature, but also slightly on density, as can be seen in Fig.~\ref{fig:opacity}. The plot shows the Rosseland mean and Planck mean opacity. If divided by a factor of $10$, the opacities in Fig.~\ref{fig:opacity} reflect the optical opacity. The opacity profile shows several bumps, which are caused by transitions in the opacity regime. For example, the decrease in opacity around $100$K is associated with the melting of ice grains, that dominate the opacity at cooler temperatures. This opacity law is applied for $\kappa_P$ and $\kappa_R$, although we point out here the Rosseland mean opacity and the Planck opacity are not the same in reality. However, for the first approximations they are quite similar. Again, the optical opacity $\kappa_\star$ is $0.1$ of the other opacities.

A change in the opacity, will ultimately lead to a change in the discs hydrodynamical structure. Higher values of the Rosseland mean opacity lead to a hotter and vertically more extended disc structure. \citet{2003A&A...410..611S} also provided opacity tables for the Rosseland mean opacity and the Planck mean opacity, however the differences between the two do not seem to be very significant, so we use here the same value for $\kappa_P$ and $\kappa_R$. The influence of different opacity laws on the disc structure will be investigated in more detail in a future paper, where we will also include a more realistic prescription for the optical opacity.

\section{Constant opacity discs}
\label{sec:constopc}

In this section we focus on the structure of discs with a constant opacity, $\kappa_P = \kappa_R = 1.0$cm$^2$/g and $\kappa_\star = 0.1$cm$^2$/g. We split the simulations into an inner disc and an outer disc so we do not have the same time step limitation due to the inner edge of the computational grid. We present in this section viscous simulations with both constant viscosity of $\nu = 10^{15}$\,cm$^2$/s and with $\alpha$ viscosity with $\alpha=0.001, 0.004, 0.008$, where $\nu = \alpha c_s^2 / \Omega$, as well as simulations of non-viscous discs. All simulations include stellar heating and the discs all feature $\Sigma_0=1000 g/cm^2$. $\Sigma (r)$ is constant for the simulations with constant viscosity, while $\Sigma (r)$ changes shape in the discs with $\alpha$ viscosity until an equilibrium is reached.

\subsection{Inner disc}

\subsubsection{constant-viscosity discs}

\begin{figure}
 \centering
 \resizebox{\hsize}{!}{\includegraphics[width=0.9\linwx]{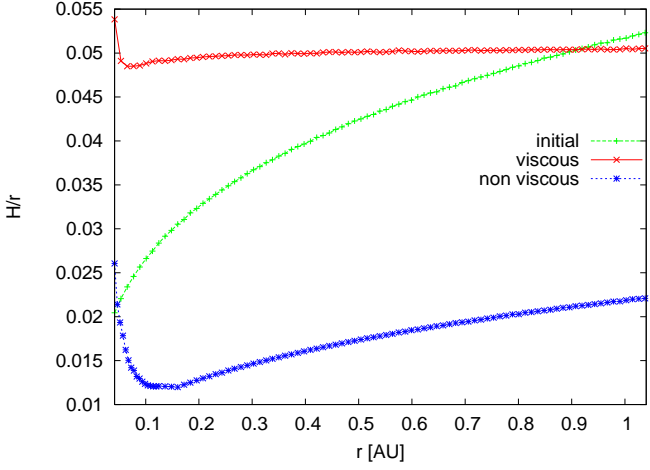}}
 \caption{Aspect ratio $H/r$ of the stellar irradiated inner disc calculated using the mid-plane values of $c_s$ for discs with constant opacity, $\kappa=1$cm$^2$/g and constant kinematic viscosity. Starting from a flared disc profile the discs are evolved until they reach thermal equilibrium.
   \label{fig:Hrinnerk1}
   }
\end{figure}

We expect the structure of the inner disc to be dominated by viscosity. The viscous heating is stronger in the inner parts of the disc compared to the stellar heating because $Q^+ \propto r^{-3}$ and $S \propto r^{-2}$ (see Equation~\ref{eq:Qplusminus}). In Fig.~\ref{fig:Hrinnerk1} we present the aspect ratio $H/r$ of discs with constant viscosity. $H$ is defined as $H=c_s / \Omega_K$, where $c_s$ is the isothermal sound speed $c_s=\sqrt{P/\rho}$ computed in the mid-plane of the disc. The discs are evolved from the initially flared structure until they reach thermal equilibrium.

The non-viscous disc (blue line in Fig.~\ref{fig:Hrinnerk1}) features a puffed-up inner rim due to stellar irradiation, while the disc behind is shadowed by the rim leading to a vertical contraction and a drop of $H/r$ in the outer parts of the simulations. In theory, if the puffed up inner rim is high enough, the whole outer disc can be shielded from stellar irradiation \citep{2002A&A...395..853D, 2004A&A...417..159D}. However, we find that beyond $0.15$AU the discs aspect ratio increases with $r$, even if the disc remains in the shadow of the inner rim. This is because the heat propagates outwards from the inner rim. Therefore the temperature drops with distance as $T = T_0 r^{-\beta}$. As the disc is in hydrostatic equilibrium, the aspect ratio increases if $\beta < 1.0$. 

For the viscous disc (red line in Fig.~\ref{fig:Hrinnerk1}) only a very small puffed up inner rim is visible, with a marginally increased $H/r$ near the inner edge of the disc. The otherwise constant aspect ratio is typical of a viscously passive disc, which we show below analytically. 

For a purely passive disc (without stellar irradiation), consider the viscous heating $Q^+$ and the radiative cooling $Q^-$ (blackbody cooling) for an annulus with size $A=2 \pi r \delta r$
\begin{equation}
\label{eq:Qplusminus}
  Q^- = 4 \pi r \delta r \sigma T^4 / \tau_{\rm eff} \quad \mathrm{and} \quad Q^+ = \frac{9}{8} \Sigma \nu \Omega_K^2 2 \pi r \delta r \ .
\end{equation}
Please be aware that the disc can cool from both sides, hence the factor $4$ in the cooling term. The effective optical depth is given by $\tau_{\rm eff}=0.5 \kappa \Sigma$. By using the $H/r$ profile for discs in hydrostatic equilibrium
\[
 T_D = \left( \frac{H}{r} \right)^2 \frac{G M_\star}{r} \frac{\mu}{\cal R}
\]
and with $\tau_{\rm eff}=0.5 \Sigma \kappa$ we can equate heating and cooling to find the aspect ratio
\begin{eqnarray}
 \frac{H}{r} &= \left( \frac{9}{32} \Sigma_0^2 \kappa \nu \frac{{\cal R}^4}{\mu^4} \frac{1}{G^3 M_\star^3 \sigma} \right)^{1/8} r^{(1-2s)/8} \ , \\
 \frac{H}{r} &= 0.051 \left( \frac{\Sigma_0}{1000 {\rm g}/{\rm {cm}}^2} \right)^{1/4} \left( \frac{r}{1 {\rm AU}} \right)^{(1-2s)/8} \nonumber \ ,
\end{eqnarray}
where $\Sigma = \Sigma_0 (r/1 {\rm AU})^{-s}$. Here we have assumed both constant opacity ($\kappa=1$cm$^2$/g) and constant kinematic viscosity. Therefore, for discs with $\Sigma = \Sigma_0 (r/1 {\rm AU})^{-0.5}$ the result is a disc with a constant aspect ratio, which matches quite well with our numerical simulations.

In total the aspect ratio of the viscous disc is higher compared to the non-viscous disc. This implies that the innermost bump of the disc in the non-viscous disc (which is due to stellar irradiation) does not influence the structure of the viscous disc, as the viscous heating dominates.

\subsubsection{$\alpha$-viscosity discs}

\begin{figure}
 \centering
 \resizebox{\hsize}{!}{\includegraphics[width=0.9\linwx]{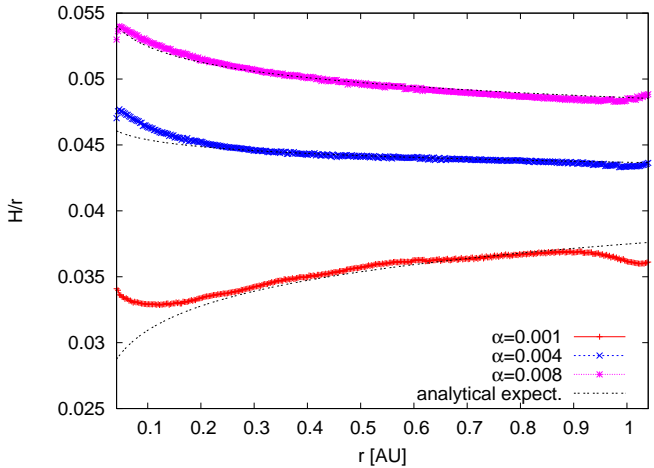}}
 \caption{Aspect ratio $H/r$ of the stellar irradiated inner disc calculated using the mid-plane values for discs with $\alpha$-viscosity and constant opacities. The black lines indicate the analytical expectations for the aspect ratio (see text).
   \label{fig:Hrinneralphak1}
   }
\end{figure}

For $\alpha$-viscosity discs we also expect the viscosity to dominate over stellar irradiation in the inner disc. The corresponding $H/r$ profiles are shown in Fig.~\ref{fig:Hrinneralphak1}. The black (dotted) lines in the plot show the analytical expectation that neglects stellar irradiation. In fact the aspect ratio of the different $\alpha$ discs vary with the value of viscosity. As the viscosity changes with $r$, the gradient of the surface density changes as well, influencing the aspect ratio.

For a purely passive disc with $\alpha$-viscosity, the aspect ratio can be computed in the same way as for a constant viscosity disc. We equate viscous heating $Q^+$ and blackbody cooling $Q^-$:
\begin{equation}
\label{eq:Qalphaplusminus}
  Q^- = 4 \pi r \delta r \sigma T^4 / \tau_{\rm eff} \quad \mathrm{and} \quad Q^+ = \frac{9}{8} \Sigma \nu \Omega_K^2 2 \pi r \delta r \ ,
\end{equation}
where $\nu = \alpha c_s H = \alpha c_s^2 / \Omega_K = \alpha H^2 \gamma \Omega_K$. With the relation of the hydrostatic equilibrium, the heating of the disc is given by
\begin{equation}
 Q^+ = \frac{9}{8} \Sigma \alpha {\cal R} \frac{\gamma}{\mu} T \Omega_K 2 \pi r \delta r \ .
\end{equation}
By equating heating and cooling, the aspect ratio of the disc can be determined:
\begin{eqnarray}
 \frac{H}{r} &= \left( \frac{9}{32} \Sigma_0^2 \gamma \frac{{\cal R}^4}{\mu^4} \frac{\kappa}{\sigma} \frac{1}{(GM_\star)^{2.5}} \alpha \right)^{1/6} r^{1/4 - s/3} \ , \\ 
 \frac{H}{r} &= 0.0572 \left( \frac{\alpha}{0.001} \right)^{1/6} \left( \frac{\Sigma_0}{1000 {\rm g}/{\rm {cm}}^2} \right)^{1/3} \left( \frac{r}{1 \rm AU} \right)^{1/4 - s/3} \nonumber \ ,
\end{eqnarray}
which clearly indicates that for a large enough gradient of the surface density $s$, the aspect ratio can decrease with $r$. This is exactly what happens for our $\alpha = 0.008$ simulation for which we measure $s=0.85$. As the inner disc is dominated by viscosity, this calculation gives an estimate for the aspect ratio of the discs that is in good agreement with our simulation (black lines in Fig.~\ref{fig:Hrinneralphak1}).

\subsection{Outer disc}

\begin{figure}
 \centering
 \resizebox{\hsize}{!}{\includegraphics[width=0.9\linwx]{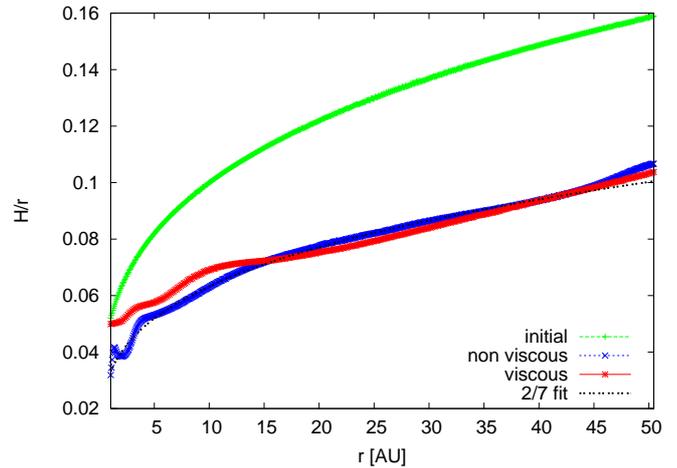}}
 \caption{Aspect ratio $H/r$ of the stellar irradiated outer disc calculated using mid-plane values for discs with constant opacity $\kappa=1$cm$^2$/g and constant viscosity. Starting from a flared disc profile the discs are evolved until they reach thermal equilibrium. These simulations differ from the ones presented in Fig.~\ref{fig:Hrinnerk1} only by the opening angle and radial extent of the disc.
   \label{fig:Hrouterk1}
   }
\end{figure}

Theoretical calculations of \citet{1997ApJ...490..368C} have shown that the outer disc is flared with $H/r \propto r^{2/7}$. This can easily be shown by equating the stellar heating with the cooling at the top of the disc. In our simulations for constant opacity we find exactly this behaviour, as can be seen in Fig.~\ref{fig:Hrouterk1} where we display the aspect ratio of discs with viscosity and without. For the simulations with viscous heating an absorbing angle of $\vartheta_{abs}=6.5^\circ$ has been used, while for the non-viscous simulations $\vartheta_{abs}=5.0^\circ$ has been used, both taken from the results of the inner disc simulations. The aspect ratio of the outer viscous disc matches perfectly with the inner disc (Fig.~\ref{fig:Hrinnerk1}), while there is some small difference for the non-viscous disc. In the non-viscous case the shadowed region behind the innermost rim (in the inner disc simulation) seems not to be captured perfectly by the continuing outer disc simulation. The non-viscous disc simulations show a slight mismatch in the $H/r$ profile between the inner disc and the outer disc. This mismatch cannot be seen in the viscous disc simulations, where $H/r$ matches very well. In reality, accretion discs are viscous, so we are confident that our code reproduces the transition between inner and outer disc very well in that case, which we will focus on in section~\ref{sec:nconstopc}. 

It can be seen clearly that both discs show a flared profile in the outer regions of the disc, which follows the predicted $r^{2/7}$ profile. In the outer regions, both disc profiles are nearly the same as stellar heating dominates over viscous heating. In the inner regions of the simulation, small differences between the viscous and non-viscous discs arise, which are due to the importance of viscous heating.

In most 2D simulations of locally isothermal discs (in $r-\phi$ plane), a constant $H/r$ profile is assumed, which corresponds to a constant opacity disc without stellar irradiation. However, in radiative discs with temperature dependent opacities the aspect ratio will depend on radius. This is discussed below.

\section{Disc structure with temperature dependent opacity}
\label{sec:nconstopc}

In realistic accretion discs, the opacity is not constant. The opacity depends on the temperature and density (Fig.~\ref{fig:opacity}). As the temperature inside an accretion disc drops from a few $1000$K in the inner parts to a few $10$K in the outer parts, we expect the opacity to vary by orders of magnitude as it was already stated in \citet{2003A&A...410..611S} and as can also be seen in Fig.~\ref{fig:opacity}. As the inner parts of the disc are dominated by viscous heating and as accretion discs are viscously driven, we focus here only on simulations including viscous heating, featuring both constant viscosity and $\alpha$ viscosity. All simulations shown here also include stellar irradiation.

\subsection{Inner disc structure}

\subsubsection{Constant viscosity}

\begin{figure}
 \centering
 \resizebox{\hsize}{!}{\includegraphics[width=0.9\linwx]{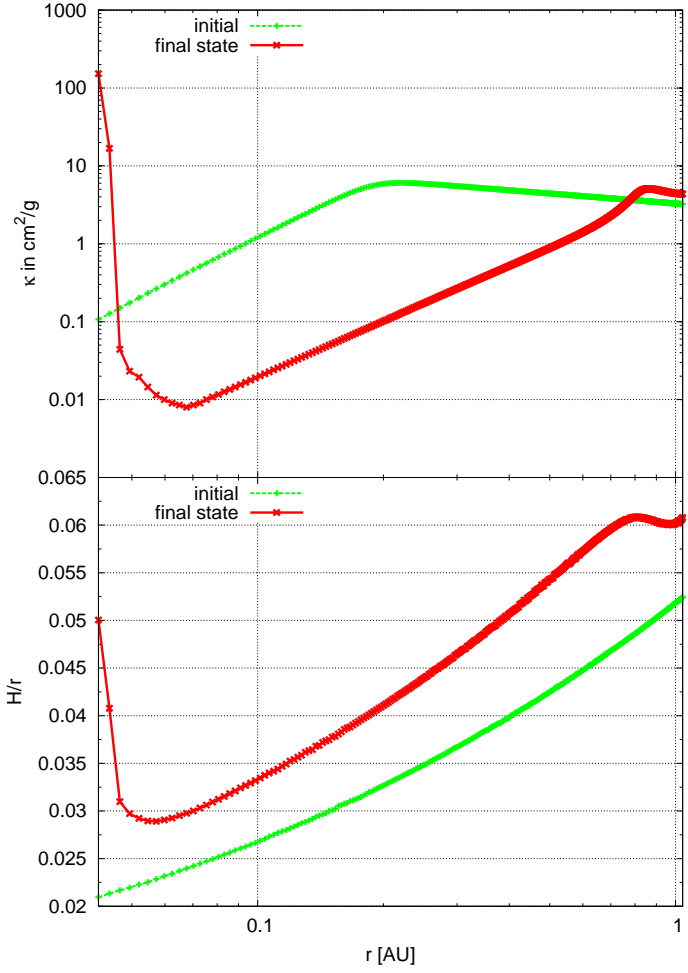}}
 \caption{Aspect ratio $H/r$ of the stellar irradiated inner disc (bottom) calculated using the mid-plane values for $\Sigma_0 =1000$g/cm$^2$ discs with non constant opacity as in Fig.~\ref{fig:opacity} and the corresponding opacity values for $\kappa_R$ (top). Starting from a flared disc profile the discs are evolved until they reach thermal equilibrium.
   \label{fig:Hrinnernck}
   }
\end{figure}

In this subsection, all simulated discs have a constant viscosity of $\nu = 10^{15}$\,cm$^2$/s. In Fig.~\ref{fig:Hrinnernck} the aspect ratio and the corresponding opacity $\kappa_R$ of the inner disc with $\Sigma_0=1000$g/cm$^2$ is displayed. Compared to the initial profile the aspect ratio increases, due to viscous heating. This effect is also be seen for a constant opacity (Fig.~\ref{fig:Hrinnerk1}). The drop behind the puffed up inner edge was not visible in the simulations with constant opacity (red line in Fig.~\ref{fig:Hrinnerk1}). As can be seen in Fig.~\ref{fig:opacity}, the large features in the opacity profile at high temperature are important for the inner disc.

At the inner edge of the grid ($r=0.0416$AU) the temperature drops from $\approx 10000$K to about $3000$K in a few grid cells, which is associated with a large drop in opacity (top in Fig.~\ref{fig:Hrinnernck}). As the opacity drops, the cooling rate increases, as it is $\propto 1/\kappa_R$. An increased cooling also means that the temperature drops even further. As the temperature is linked to $H/r$, a drop in temperature translates into a drop in aspect ratio. The drop of $H/r$ behind the puffed up inner edge is directly related to a drop in opacity, as we show in Appendix~\ref{ap:Hrdrop}.

If the inner edge were be shifted towards smaller radii, the height of the puffed up inner region would not increase indefinitely, as the opacity does not continue to increase, but rather levels off at high temperatures (see Fig.~\ref{fig:opacity}). 

As soon as the temperature drops to lower values, the opacity increases again (Fig.~\ref{fig:opacity} and top of Fig.~\ref{fig:Hrinnernck}), which in turn increases $H/r$. At around $r=0.83$AU we observe another bump in the $H/r$ profile. Here the temperature crosses the $\approx 1000$K bump in the opacity profile, which leads to a decrease of $H/r$ for $r>0.83$AU. 

We can clearly relate the transition regions of the opacity to transition regions of the $H/r$ profile of the disc. These changes in the disc's profile are of crucial importance for irradiated discs. The bump at $r=0.83$AU is higher than the bump at the inner edge, indicating that stellar irradiation will be absorbed at a larger height from mid-plane, which means in turn that the disc will only be heated above a larger angle $\vartheta_{abs}$. As the inner region of the disc is dominated by viscous heating, the height and location of the bumps in the disc will change with disc mass and the amount of viscous heating. In contrast, stellar irradiation is relatively unimportant in the inner disc and does not change significantly with disc mass.

\subsubsection{$\alpha$-viscosity}

\begin{figure}
 \centering
 \resizebox{\hsize}{!}{\includegraphics[width=0.9\linwx]{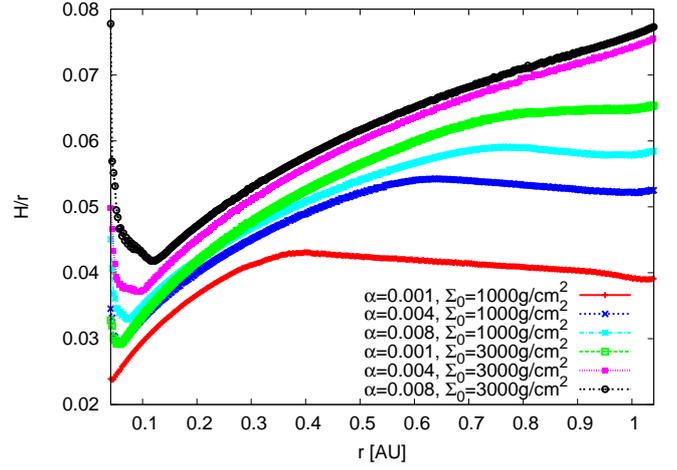}}
 \caption{Aspect ratio $H/r$ of the inner disc (with viscous and stellar irradiation) calculated using the mid-plane values for discs with $\alpha$-viscosity and non constant opacity as in Fig.~\ref{fig:opacity}. Starting from a flared disc profile the discs are evolved until they reach thermal equilibrium.
   \label{fig:Hralphainnernck}
   }
\end{figure}

The amount of viscous heating has a crucial influence on the disc structure in the interior of the disc. We now focus on discs with different $\alpha$ viscosities, with $\nu = \alpha c_s^2 / \Omega_K$, where we use $c_s$ in the mid-plane of the disc. The results of simulations of the inner disc with different $\alpha$ and $\Sigma_0$ values are shown in Fig.~\ref{fig:Hralphainnernck}.

As expected the aspect ratio of the disc increases with increasing $\alpha$ parameter due to the increased viscous heating. Also, the aspect ratio increases as we increase $\Sigma_0$, as more mass inside the disc results in a larger heating of the disc. This has the effect that the peaks in the $H/r$ profile are shifted towards larger radii, as the temperature increases and therefore the transitions in the opacity are reached at larger distances from the star.

The simulation of constant viscosity (with $\Sigma_0=1000$g/cm$^2$ in Fig.~\ref{fig:Hrinnernck}) most closely resembles the $\alpha=0.004$ simulation. However, the value that matches best might be around $\alpha \approx 0.0055$.

With increasing disc mass and viscosity an increase of the height of the peaks in the discs is visible. These bumps in the $H/r$ profile are not directly related to the absorption angle $\vartheta_{abs}$ of the disc, but it is safe to say that a higher bump in the disc results in a larger absorption angle with important effects on the structure of the outer disc, as discussed below.

\subsection{Outer disc}

\subsubsection{Constant viscosity}

\begin{figure}
 \centering
 \resizebox{\hsize}{!}{\includegraphics[width=0.9\linwx]{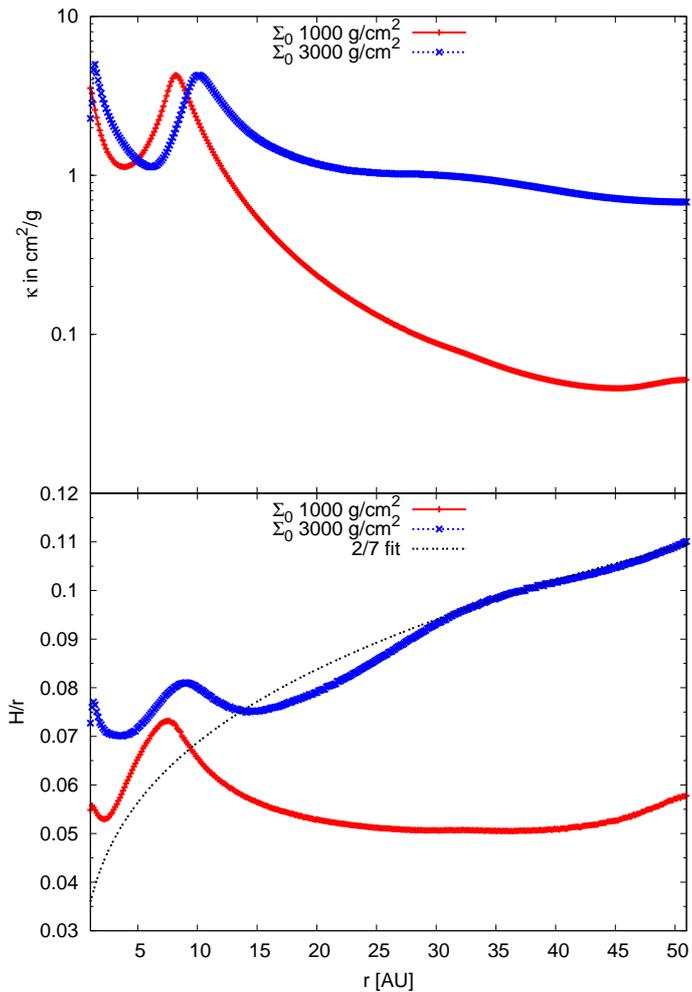}}
 \caption{Aspect ratio $H/r$ of the stellar irradiated outer disc (bottom) calculated using the mid-plane values for discs with constant viscosity and non constant opacity as in Fig.~\ref{fig:opacity} and the corresponding opacity values for $\kappa_R$ (top). Starting from a flared disc profile the discs are evolved until they reach thermal equilibrium. The different colours indicate different $\Sigma_0$-values. The red line represents the outer disc simulation corresponding to the inner disc simulation shown in Fig.~\ref{fig:Hrinnernck}.
   \label{fig:Hrouternck}
   }
\end{figure}

We have shown in Section~\ref{sec:constopc} that for discs with constant opacity the discs are flared with $H/r \propto r^{2/7}$ in agreement with \citet{1997ApJ...490..368C}. However, for non constant opacities we expect another feature in the disc's structure as the temperature crosses $100$K (Fig.~\ref{fig:opacity}). The aspect ratio of the outer disc structure and the corresponding opacity $\kappa_R$ for two different surface density values are presented in Fig.~\ref{fig:Hrouternck}. 

At $r=1.04$AU the aspect ratio of the $\Sigma_0=1000$g/cm$^2$ outer disc does not exactly match the aspect ratio computed for the inner disc (red line in Fig.~\ref{fig:Hrinnernck}). In this region the disc is shadowed from direct stellar irradiation and any radial transfer of energy relies on the re-radiated radiative flux. The outer simulation does not include a source of the re-radiated flux at the inner boundary, leading to the small difference in the inner disc. 

The disc with $\Sigma_0=1000$g/cm$^2$ features another bump in the $H/r$ profile at $r\approx 7.3$AU. At this point in the disc the temperature crosses $T\approx 100$K, which is the transition region for melted ice grains to solid ice grains. The increase of $H/r$ for $r<7.3$AU is related to the increase of opacity for $200$K $> T > 100$K, which can clearly be seen in the top of Fig.~\ref{fig:Hrouternck}. The minima and maxima of opacity and $H/r$ are not exactly at the same location, but are slightly shifted. However, a clear trend is still visible. Interestingly, for $r>7.3$AU the $H/r$ profile monotonically decreases and the profile follows that of an non-irradiated disc due to the self-shadowing of the disc.

The fact that the disc is not flared in the outer parts of the disc is a combination of the value of the surface density and of the opacity. In the constant opacity scenario, the outer parts of the disc were flared (Fig.~\ref{fig:Hrouterk1}) for the $\Sigma_0=1000$g/cm$^2$ disc. However, in reality for lower temperatures ($T<100$K), the opacity drops to very small values (see top in Fig.~\ref{fig:Hrouternck}). Please keep in mind that $\kappa_\star=0.1 \kappa_R$. As the opacity drops, so does the optical depth $\tau=\rho \kappa_\star \Delta r$ of each grid cell, which means that less and less stellar irradiation is absorbed in the outer regions of the disc, as the absorption is proportional to $\tau$ (eq.~\ref{eq:stellarthin}).

In order to keep the disc flared, the upper layers in the outer parts of the disc have to absorb stellar irradiation. As the opacity drops in this region of the disc, an increase in density would keep the disc flared, as $\tau = \kappa_\star \rho \Delta r$. Indeed, this can be seen in Fig.~\ref{fig:Hrouternck}, where the blue line indicates a disc with $\Sigma_0=3000$g/cm$^2$. The outer parts are flared and follow the $2/7$th profile. Also interesting to note is that the bumps in the inner region of the disc in the $H/r$ profile are shifted to larger distances from the star compared to the lower mass disc (red in Fig.~\ref{fig:Hrouternck}). This is due to the fact that the higher mass disc produces more viscous heating and is therefore warmer. We compare the disc with $\Sigma_0=3000$g/cm$^2$ in great detail with a disc that is only heated through viscosity in Section~\ref{sec:discs}.

\subsubsection{$\alpha$-viscosity}

\begin{figure}
 \centering
 \resizebox{\hsize}{!}{\includegraphics[width=0.9\linwx]{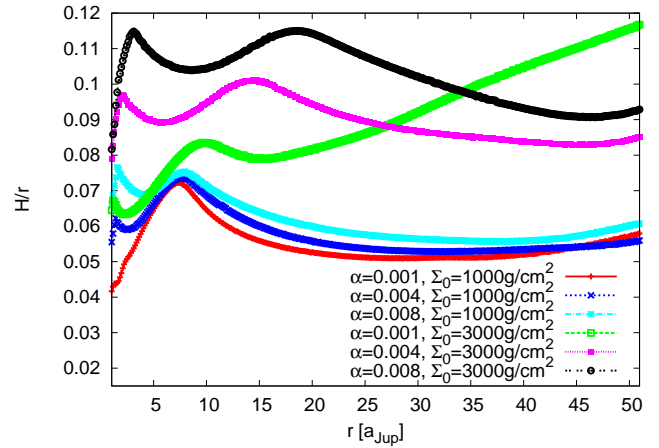}}
 \caption{Aspect ratio $H/r$ of the outer disc (with $\alpha$-viscosity and stellar irradiation) calculated using the mid-plane values for discs with non constant opacity as in Fig.~\ref{fig:opacity}. Starting from a flared disc profile the discs are evolved until they reach thermal equilibrium. These simulations differ from the ones presented in Fig.~\ref{fig:Hralphainnernck} only by the opening angle and radial extent of the disc.
   \label{fig:Hralphaouternck}
   }
\end{figure}

The bumps of the inner disc can shield the outer disc from stellar irradiation implying the outer disc structure is influenced by the inner disc structure, which, in turn, is determined by viscosity and disc mass. In Fig.~\ref{fig:Hralphaouternck} the aspect ratio for discs with different $\alpha$-viscosity and disc mass are displayed.

The $\Sigma_0=1000$g/cm$^2$ discs are not flared for any viscosity. The outer disc is just too thin to absorb stellar irradiation effectively and viscosity is negligible there, as was seen the constant viscosity disc (Fig.~\ref{fig:Hrouternck}). However, we do see the innermost bump in the disc rise as the viscosity increases.

For the $\Sigma_0=3000$g/cm$^2$ disc with $\alpha=0.001$ the equilibrium disc structure features a flared disc profile, which is quite similar to that of a disc with constant opacity (Fig.~\ref{fig:Hrouternck}). However, for $\alpha>0.004$, the discs are not flared any more. As the viscosity increases the inner bumps of the disc structure and block stellar irradiation from going through to the outer parts of the disc. This effect increases with increasing viscosity. 

In order to have a disc flared in the outer regions, the disc must not only have the right amount of mass, but also the viscosity of the disc should not be too large, as the viscously dominated inner disc can block stellar irradiation from the outer disc. Therefore, simulations of the inner disc are always needed to determine the structure of the outer disc correctly. Whether a disc is shadowed or not, is determined through viscosity and disc mass.

\subsection{Passive disc}
\label{sec:discs}

We now focus on the difference between stellar irradiated discs and discs without stellar irradiation. For this purpose we refer to the case of constant viscosity $\nu$ and $\Sigma_0=3000$g/cm$^2$, as it was already shown in Fig.~\ref{fig:Hrouternck}.

\begin{figure}
 \centering
 \resizebox{\hsize}{!}{\includegraphics[width=0.9\linwx]{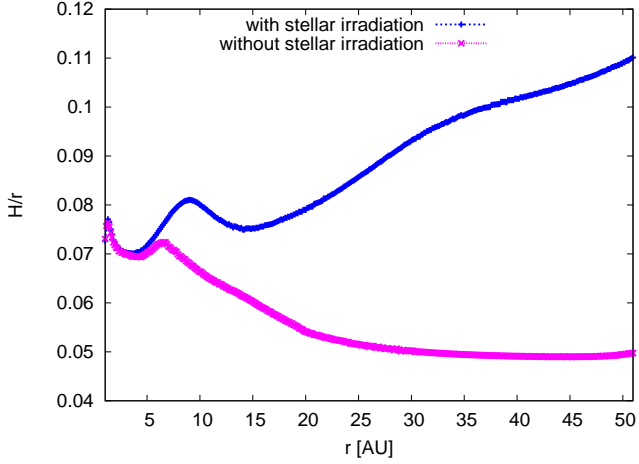}}
 \caption{Aspect ratio $H/r$ of the outer disc calculated using the mid-plane values for discs with constant viscosity and non constant opacity as in Fig.~\ref{fig:opacity}. Starting from a flared disc profile the discs are evolved until they reach thermal equilibrium. The plot features one disc with and one without stellar heating; both discs have a constant viscosity.
   \label{fig:Hroutercomp}
   }
\end{figure}

In Fig.~\ref{fig:Hroutercomp} the two aspect ratios of the disc with $\Sigma_0=3000$g/cm$^2$ are displayed. In the inner parts, the aspect ratios for both discs are nearly identical, as this region is dominated by viscous heating and not by stellar irradiation. Beyond $\approx 20$AU the stellar irradiated disc is flared, while the disc without stellar irradiation collapses to small $H/r$ at large radii. The $100$K bump at $r \approx 9$AU is more prominent in the stellar irradiated disc and also shifted a little bit towards larger distances compared to the non-stellar irradiated disc, as the disc receives more heat which shifts the bump.

\begin{figure}
 \centering
 \resizebox{\hsize}{!}{\includegraphics[width=0.9\linwx]{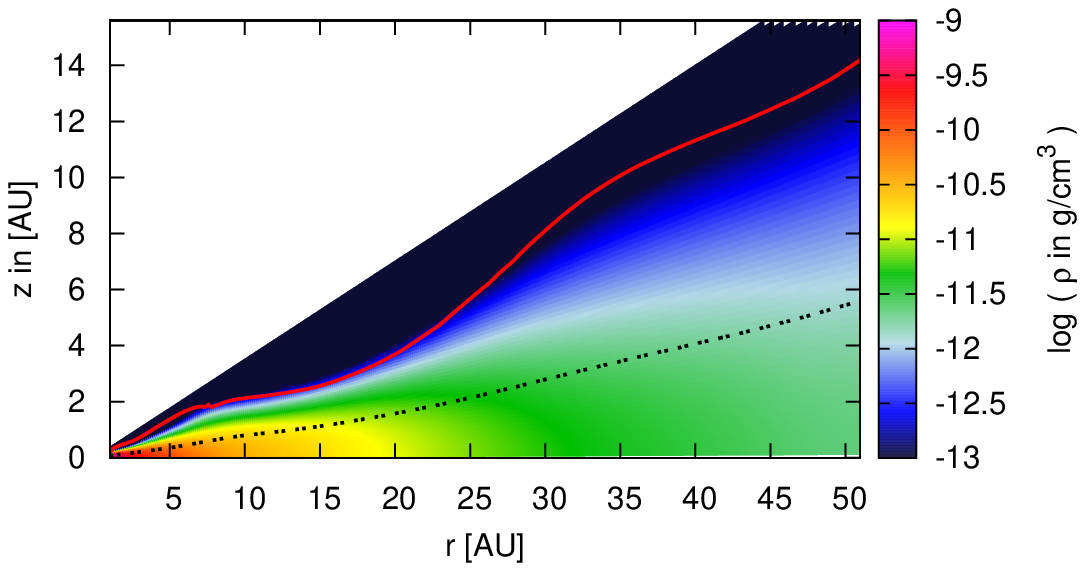}}
 \resizebox{\hsize}{!}{\includegraphics[width=0.9\linwx]{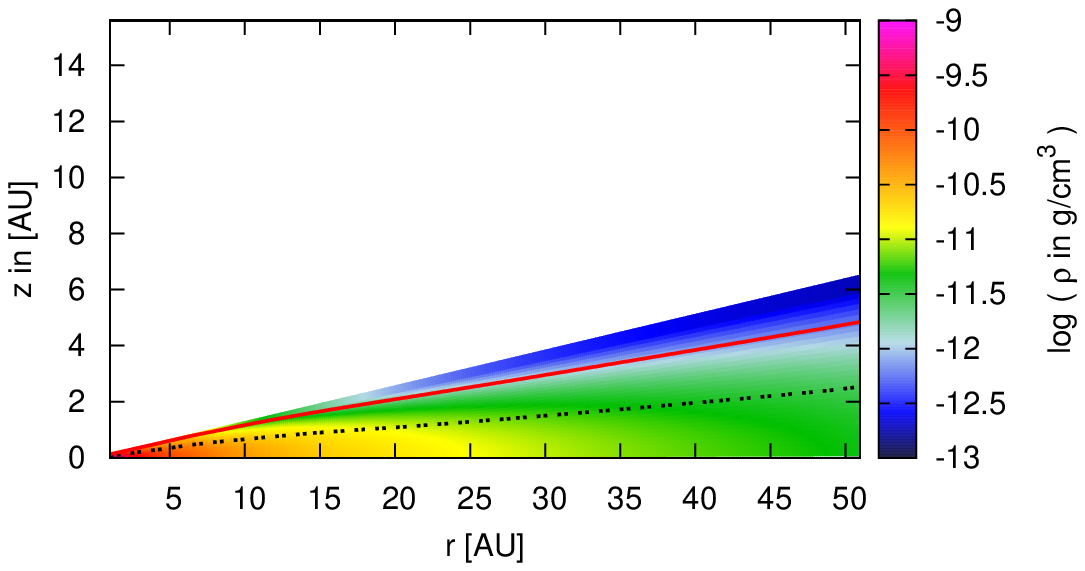}}
 \caption{Density map (in log-scale) of discs with stellar heating (top) and without stellar heating (bottom). Both discs undergo viscous heating and have $\Sigma_0=3000$g/cm$^2$. The black dotted line indicates $H(r)$ and the red solid line represents the $\tau_l=1.0$ line integrated from the top of the disc.
   \label{fig:Rhocomp}
   }
\end{figure}

The 2D density map of the discs is shown in Fig.~\ref{fig:Rhocomp}. The density profile of the disc without stellar heating does not extend to the same height, because we only use $83^\circ \leq \theta \leq 90^\circ$ for this disc. We reduced the vertical extent of the grid in this case because of the smaller disc thickness. In the disc with stellar irradiation, on the other hand, the opening angle of the grid is larger, as we need the disc to absorb the stellar irradiation in the top layers.

In Fig.~\ref{fig:Rhocomp} the $H(r)$ line is indicated in black. In both cases the $H(r)$ line rises, however, only for the stellar irradiated disc does $H/r$ increase (Fig.~\ref{fig:Hroutercomp}). On the other hand, the bumps, which are clearly visible in the $H/r$ profiles are not visible in the $H(r)$ profile, because the increase of $H(r)$ is too strong to notice the small bumps.

Of interest is also the red line in Fig.~\ref{fig:Rhocomp}, which refers to the $\tau_l=1.0$ line integrated from the top (infinity) of the disc. We define in this case 
\begin{equation}
 \tau_l (z) = \int_{\infty}^z \rho \kappa_R dz' \ ,
\end{equation}
where $z$ is the vertical thickness of the disc and the integration is performed on lines of constant radius $r$. This line is roughly connected to the flux $\mathbf{F}$ (eq.~\ref{eq:raddif}) of the disc and therefore gives a rough estimate of the location of the photosphere where the disc is cooling. One can clearly see the bumps in the $\tau_l=1.0$ line for the stellar irradiated disc. These bumps are at the same location as the ones in the $H/r$ profile, which are not visible in this figure. As $\tau$ depends on $\kappa_R$ and $\rho$, it is no surprise to see a drop of density just above the $\tau_l=1.0$ line around $r \approx 15.6$AU. The bumps of the $\tau_l$ line can not be seen in the non-irradiated disc.
This line just levels off for large distances to the star, where the disc must contract from its initial state in order to maintain its energy loss to the upper and lower layers.

\begin{figure}
 \centering
 \resizebox{\hsize}{!}{\includegraphics[width=0.9\linwx]{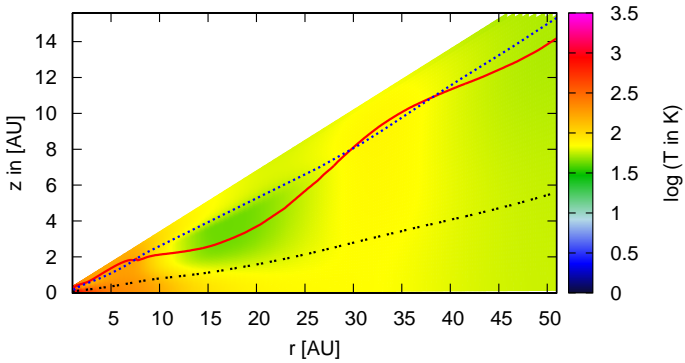}}
 \caption{Temperature map (log-scale) of a disc with stellar and viscous heating. Indicated in red is the $\tau_l=1.0$ line, blue the $\tau_\star=1.0$ line and in black the $H(r)$ line. The disc has $\Sigma_0=3000$g/cm$^2$.
   \label{fig:Temp2D}
   }
\end{figure}

One of the most important indicators of the disc structure is the temperature of the disc, as the temperature is related to the entropy. Due to the stellar irradiation we expect the upper layers of the disc to be heated from the star, while the mid-plane regions far away from the star are heated vertically from the upper layers. The temperature in mid-plane might therefore be slightly smaller compared to the top layers. The temperature is displayed in Fig.~\ref{fig:Temp2D}. Again we over plot the line for $\tau_l=1.0$ (red), $H(r)$ (black) and additionally in blue the $\tau_\star=1.0$ line, which we define as
\begin{equation}
  \tau_\star (r) = \int_0^r \rho \kappa_\star dr' \ ,
\end{equation}
where the integration is performed along lines of constant $\theta$. The $\tau_\star$ line represents where in the disc the stellar heating is absorbed effectively. This line does not match the $\tau_l$ line at $10$AU$<r<25$AU, where $\tau_\star > \tau_l$. This means the stellar heating of the disc is effective higher above the mid-plane compared to the cooling, which results in a higher temperature at the upper layers of the disc, as can be seen clearly in Fig.~\ref{fig:Temp2D}.

The upper layers of the disc are heated by stellar irradiation, while the disc's interior is also heated through viscosity. However, the heating effect of viscosity is diminished in the upper layers of the disc, as the density is much lower compared to the mid-plane of the disc (see Fig.~\ref{fig:Rhocomp}). In the region of the bump in optical depth ($10.4$AU $< r < 23.4$AU), the effect of stellar heating is reduced as the opacity is reduced, which leads to less absorption of stellar irradiation. In this region the upper layers of the disc are much hotter compared the regions of the disc right below. This temperature inversion in vertical direction of the disc could have interesting implications to planets on inclined orbits in this part of the disc, as the migration speed is dependent on the gradient of temperature. 

In the outer regions of the disc, viscosity is negligible and the disc is heated solely by stellar irradiation. The heat is then transferred from the upper regions to mid-plane, which causes the temperature to show only little vertical dependence.

Observations indicate that protoplanetary discs have an outwards flared profile, which is consistent with our stellar irradiated model. The differences in the disc structure between the two models are dramatic, and we therefore recommend using stellar irradiated disc models for the outer parts of the disc. However, as we have seen in Fig.~\ref{fig:Hrouternck} and Fig.~\ref{fig:Hralphaouternck}, stellar irradiated discs can be self shadowed if the disc mass is small or the viscosity large. In both cases the inner bumps of the discs are so large that they block stellar irradiation from propagating to the outer parts of the disc. For low disc mass models, passive discs follow the same $H/r$ profile as stellar irradiated discs, because the discs are optically thin and can not maintain the flared disc profile.

\section{Implications to planet migration}
\label{sec:migration}

In order to estimate what is the difference between the torque acting on planets in discs with and without stellar irradiation, we apply the torque formula of \citet{2011MNRAS.410..293P} to the discs described in Section~\ref{sec:discs}. These discs feature $\Sigma_0=3000$g/cm$^2$, a constant viscosity $\nu$ and the opacity law in Fig.~\ref{fig:opacity}. The formula captures the behaviour of the torque caused by Lindblad resonances and horseshoe drag on low-mass planets embedded in gaseous discs in the presence of viscous and thermal diffusion. This formula gives the best match to the full 3D radiative simulations of \citet{2011A&A...536A..77B}. We do not simulate planets embedded in stellar irradiated discs in this work, as it is beyond the scope of the paper at this point.

The formula of \citet{2011MNRAS.410..293P} is very complex and features many details, so that we do not cite every step of the formula at this point. The formula is also additionally displayed in the Appendix of \citet{2011A&A...536A..77B}. However, we want to point out a few key items of the formula. The total torque acting on an embedded planet is a composition of its Lindblad torque and its corotation torque:
\begin{equation}
 \Gamma_{tot} = \Gamma_L + \Gamma_C
\end{equation}
The Lindblad torque depends on the gradients of temperature $T \propto r^{-\beta}$ and surface density $\Sigma \propto r^{-s}$. It is given in \citet{2011MNRAS.410..293P} by
\begin{equation}
 \gamma \Gamma_L / \Gamma_0 = -2.5 - 1.7 \beta + 0.1 s \quad \mathrm{and} \quad \Gamma_0 = \left(\frac{q}{h}\right)^2 \Sigma_P r_p^4 \Omega_P^2 \ ,
\end{equation}
where $q$ is the mass ratio between planet and star, $h=H/r$, $\Sigma_P$ the surface density of the disc at the planets location and $r_P$ the distance of the planet to the host star. One can clearly see that a change in the gradient of temperature influences the Lindblad torque. The same applies to the corotation torque, which is strongly dependent on the gradient of entropy, $S \propto r^{-\xi}$, with $\xi = \beta - (\gamma - 1.0) s$. The largest contribution of the corotation torque arises from the entropy related horseshoe drag, which is given by
\begin{equation}
  \gamma \Gamma_{hs,ent} / \Gamma_0 = 7.9 \frac{\xi}{\gamma} \ .
\end{equation}

The aspect ratio $H/r$ of the disc with and without stellar irradiation changes from flared to non flared in the outer parts of the disc. This change is related to a change in temperature and the local temperature gradient. In Fig.~\ref{fig:Temp2D} the outer parts of the disc seem to have a very small radial temperature gradient that reduces the effect on the entropy related horseshoe drag. In the disc without stellar irradiation, on the other hand, planets can migrate outwards to quite large distances, depending on the disc mass \citep{2011A&A...536A..77B}. We therefore expect that the zero-torque radius for planets will be at a smaller distance to the star for stellar irradiated discs compared to discs with only viscous heating, as the temperature gradient is larger in the latter case.

\begin{figure}
 \centering
 \resizebox{\hsize}{!}{\includegraphics[width=0.9\linwx]{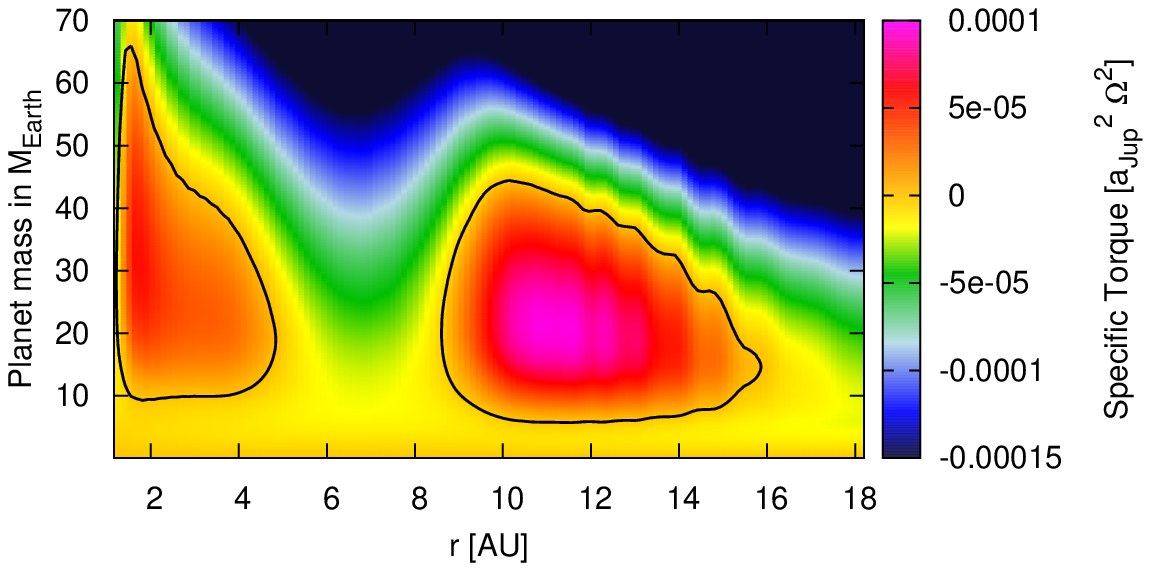}}
 \resizebox{\hsize}{!}{\includegraphics[width=0.9\linwx]{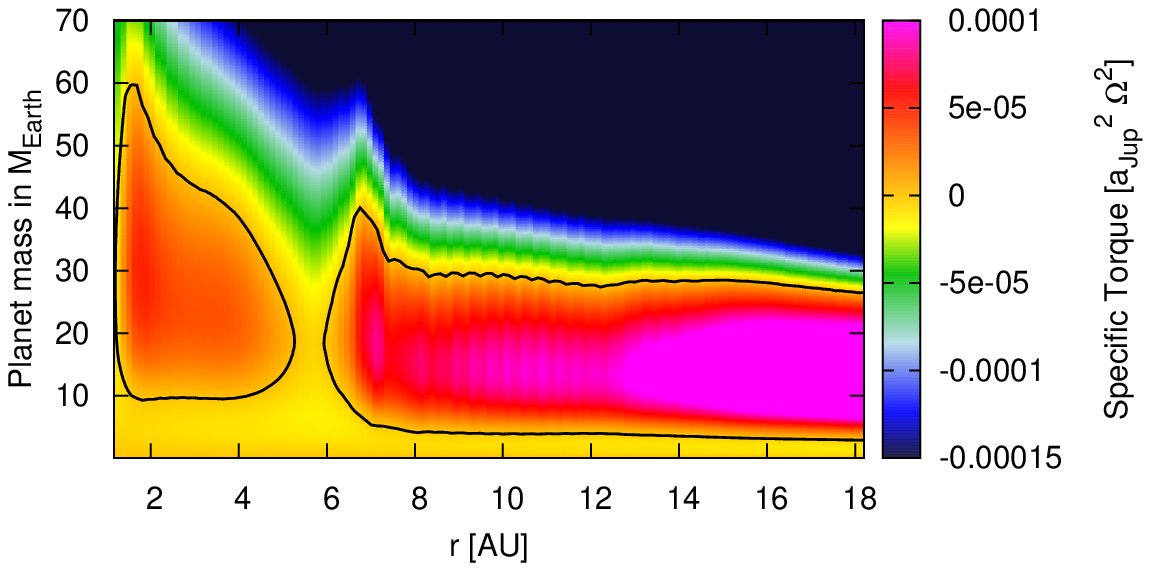}}
 \caption{Torque acting on planets with different masses using the formula of \citet{2011MNRAS.410..293P}. The top plot features a disc with stellar irradiation and viscous heating (constant $\nu$) and the bottom plot a disc with just viscous heating. The colour coding of the torque has been cut off at $\Gamma_{tot}=-0.00015$, so that everything that is black in the figure features a large negative torque. The black lines encircle the regions of outward migration.
   \label{fig:Torque}
   }
\end{figure}

In Fig.~\ref{fig:Torque} the torque acting on planets with different masses in a stellar irradiated disc and a non stellar irradiated disc is displayed. The black lines encircle the regions of outward migration. If a planet is inside the black circles it will migrate outwards, but it migrates inwards for the rest of the positions in the diagram. At the left side of the black circles, a planet would face an unstable torque equilibrium, as the planets would migrate away from the line in both directions, while at the right side of the circles the planet are in a stable equilibrium as they migrate towards the line from both directions. 

We only focus here on planets with a few earth masses, as larger planets open gaps in discs and migrate inwards in type-II-migration \citep{  1986ApJ...307..395L,1986ApJ...309..846L,2006Icar..181..587C,2007MNRAS.377.1324C}. We only display the torque up to $r_P = 18.2$AU as the torque acting on planets in the stellar irradiated disc is negative for larger distances to the star. On the other hand, the torque is still positive at that point in the disc with only viscous heating. In fact the torque is still positive up to $r\approx 47$AU. One can also clearly distinguish two different regions of positive torque, indicating areas of outward migration. Finally, the region of outward migration starts at a larger distance to the star for the stellar irradiated disc. 

For both disc types the inner region ($1.56 < r < 4.7$AU) of positive torque is nearly identical, due to the fact that the disc structure is similar in the inner parts of the disc, as the disc there is dominated by viscous heating and not stellar irradiation. However, as soon as the planetary mass increases above $40M_{Earth}$ the region of outward migration becomes smaller with increasing planetary mass, and the planets migrate inward for all distances for $M_P > 60 M_{Earth}$.

Planets can migrate outwards as long as the libration time-scale is comparable to the cooling time-scale \citep{2008ApJ...672.1054B}. For large distances to the central star \citep{2011A&A...536A..77B} the density is lower in the disc, leading to a shorter cooling time-scale that prevents outward migration. 

As the planetary masses increase ($M_P > 35 M_{Earth}$) they can start to open up partial gaps inside the disc. The positive corotation torque originates from a region very close to the planet \citep{2008A&A...487L...9K, 2009A&A...506..971K}, which is then depleted as the gap starts to form, thus preventing outward migration.

The regions of outward migration in the disc relate to the $H/r$-profile. As can be seen in Fig.~\ref{fig:Hroutercomp}, whenever $H/r$ decreases in the disc, the planet migrates outwards. When $H/r$ increases the planet migrates inwards. This means, whenever we have a change in the $H/r$ profile, we change the direction of migration. As the bumps and dips in the $H/r$ profile are influenced by the opacity transitions, the opacity determines the migration. As the transitions of opacity are shifted away from the star with higher viscosity and larger surface density, so is the zero-torque radii in the disc.

To clarify the statement above, keep in mind that the torque depends on the surface density, temperature and entropy gradient, $\Gamma = f(s,\beta,\xi)$.
\begin{eqnarray}
 &\frac{H}{r} \propto r^{-b} \quad \Leftrightarrow \quad \left( \frac{H}{r} \right)^2 \propto r^{-2b} \quad \Leftrightarrow \quad \left( \frac{H}{r} \right)^2 \frac{GM_\star}{r} \frac{\mu}{R} \propto r^{-2b-1} \nonumber \\
 &\Leftrightarrow \quad T(r) \propto r^{-1-2b} = r^{-\beta} \ ,
\end{eqnarray}
which indicates if the disc shows a drop off in $H/r$, the temperature gradient increases, which therefore increases the entropy gradient as $\xi = \beta - (\gamma - 1.0) s$. Therefore the contribution of the entropy related corotation torque is stronger if $H/r$ decreases. In the flaring part of the disc, where $H/r \propto r^{2/7}$, the entropy gradient is smaller as $b=-2/7$, which leads to $\beta=3/7$, which reduces the entropy gradient $\xi$, resulting in a negative torque.

On the other hand, it seems that for planet masses with $M_P<7 M_{Earth}$ the torque is always negative and that planets migrate inwards at all radii. The reduction of the positive torque for low mass planets is due to the fact that the horseshoe region narrows, thus the horseshoe drag becoming less pronounced, resulting in a smaller net-torque acting on the planet.

For these small planetary masses the inward migration speed is about the same for the passive and stellar irradiated disc. It also seems that the minimum mass a planet needs to sustain outward migration is increased in the case of a stellar irradiated disc, compared to the non stellar irradiated case. The reason for this is a slight change in the entropy gradient of the disc, which is steeper in the non stellar irradiated disc. However, we believe that even smaller mass planets can migrate outwards as the torque formula shows several issues that should be kept in mind:
\begin{itemize}
 \item The formula by \citet{2011MNRAS.410..293P} is dependent on the smoothing of the planetary potential, which influences the torque. A change in the smoothing changes the torque acting on the planet (see Appendix in \citet{2011A&A...536A..77B}). For Fig.~\ref{fig:Torque} we used a smoothing of $0.4$.
 \item The formula shows some differences to full 3D simulations \citep{2011A&A...536A..77B}, as it was derived from 2D simulations.
 \item 3D radiation-hydrodynamical simulations have shown that small mass planets with $5M_{Earth}$ can still migrate outwards \citep{2009A&A...506..971K}.
\end{itemize}
In order to verify these assumptions, high resolution 3D simulations with embedded small mass planets must be done, as the minimal mass for outward migration has not been investigated with enough detail. We will address this issue in a further paper.

However, the torque formula gives an estimate of the strength of migration in the disc. The difference in the range of outward migration for the two disc types does not change the implication of the existence of the zero-torque radius. At zero-torque distance, planetary embryos and protoplanets can gather and collide and then form larger objects. This is of crucial importance for the growth of massive planets, where a core of a few earth masses needs time to accrete gas in order to form a gas giant planet \citep{1996Icar..124...62P}.

\section{Summary}
\label{sec:summary}

We have investigated the influence of opacity and stellar irradiation on the structure of protoplanetary accretion discs. Utilizing our calculated disc structures we have estimated the torque on embedded planets in discs with and without stellar irradiation by using the theoretical torque formula from \citet{2011MNRAS.410..293P}.

Before investigating the influence of opacity on the disc structure, we compared non-viscous and viscous discs in the inner and outer regions of the disc for a constant opacity. In the non-viscous disc, a puffed-up inner rim shields the first part of the disc from stellar irradiation, while the outer part of the disc is flared with $H/r \propto r^{2/7}$, as was shown in theoretical calculations by \citet{1997ApJ...490..368C}. In the viscous case, in the inner part of the disc, viscous heating dominates over stellar irradiation. By equating viscous heating with radiative cooling, we have shown that the aspect ratio $H/r$ is constant until stellar heating dominates. This was confirmed by our simulations.

We follow the opacity law of \citet{1994ApJ...427..987B} for our non constant opacity discs. As the opacity changes due to transitions in the opacity law, so does the disc structure. We have seen that the bumps and dips in the opacity law reflect the bums and dips in the discs structure for stellar irradiated and non irradiated discs. If the disc is more massive, the disc produces more viscous heating, which results in a higher temperature at larger distances. But as the temperature determines the opacity, the bumps in the disc are moved outwards with increasing temperature.

The outer parts of the disc ($r>7.8$AU) can be flared if the disc mass is high enough. The effect of the disc mass was not visible in the constant opacity simulations, as a constant opacity leads to a larger optical depth in the upper, less dense regions of the disc. The optical depth is crucial in determining how much stellar irradiation is absorbed. If the disc is optically thin, the stellar irradiation just passes through the disc without any heating effects. Therefore the disc absorbs less heat and can not sustain the flared profile. A minimum mass inside the disc is needed in order to keep the discs flared.

For increasing viscosity, the aspect ratios of the discs increase. This also means that the bumps due to the opacity in the disc become higher. A higher bump in the disc leads to more absorbed stellar irradiation in the region of the bump. If the bump is high enough, it prevents stellar irradiation from reaching the outer disc, resulting in a non-flared disc. Very high viscosity therefore leads to self-shadowed discs (Fig.~\ref{fig:Hralphaouternck}).

In the inner parts of the disc, where viscous heating dominates, the disc structures for discs with and without stellar irradiation are similar. In the outer regions, the stellar irradiated disc can be flared, which is not the case for the non-stellar irradiated disc. In the outer parts of the stellar irradiated disc, where the viscous heating is negligible, the disc shows a small dependence in $T(z)$, as the heat is transported from the upper layers towards mid-plane.

To estimate the torque acting on planets in discs with stellar irradiation, we apply the torque formula by \citet{2011MNRAS.410..293P}, where we expect the different disc structure in the outer parts to result in different migration scenarios. The discs feature two regions of outward migration. One in the inner part of the disc ($1.56<r<4.7$AU) that is nearly identical for the discs with and without stellar irradiation, as viscous heating dominates in that region of the disc. The second region of outward migration is completely different for the two discs. It is much smaller for the case of stellar irradiation, as the very outer parts of the disc have nearly a constant temperature, which favours inward migration. In the case of only viscous heating, outward migration continues to much larger distances from the star ($r \approx 47$AU).

However, important here is that the disc features regions of outward migration for different planetary masses. At the corners of these regions (zero-torque radius), planetary embryos can accumulate and grow until the cores are large enough to accrete gas and form gas giant planets. We note that the results from this torque formula are only an estimate and real 3D simulations with embedded planets need to be done, especially for low mass planets, as they all would migrate towards the star according to the torque formula. 

As the differences between stellar irradiated discs and passive discs are dramatic in the outer regions of the disc, we recommend using stellar irradiated disc models for high mass discs. Low mass discs are self-shadowed, so that the structure is well captured by passive disc models as well.

\begin{acknowledgements}

B. Bitsch and A. Morbidelli have been sponsored through the Helmholtz Alliance {\it Planetary Evolution and Life}. W. Kley has received support through the German Research Foundation (DFG) through grant KL 650/11 within the Collaborative Research Group FOR 759: {\it The formation of Planets: The Critical First Growth Phase}, and gratefully acknowledges the very kind hospitality of the Observatoire de Cote d'Azur. I. Dobbs-Dixon is supported by the Carl Sagan Postdoctoral program. The calculations were performed on systems of the Computer centre of the University of T\"ubingen (ZDV) and systems  operated by the ZDV on behalf of bwGRiD, the grid of the Baden  W\"urttemberg state. Finally, we gratefully acknowledge the helpful and constructive comments of an anonymous referee.

\end{acknowledgements}

\appendix
\section{Stellar heating}
\label{ap:heat}

\subsection{Calculation of stellar heating in the code}

To compute the amount of stellar heating that is deposited in a grid cell, we compute the difference between what arrives and what leaves a grid cell:
\begin{eqnarray}
 F_\star e^{-\tau_i} - F_\star e^{-\tau_{i+1}} & = & F_\star e^{-\tau_i} \left(1 - \frac{e^{-\tau_{i+1}}}{e^{-\tau_i}} \right) \\
 = F_\star e^{-\tau_i} \left(1 - e^{-(\tau_{i+1} - \tau_i)} \right) & = & F_\star e^{-\tau_i} \left(1 - e^{-\rho_i \kappa_i \Delta r} \right) \ ,
\end{eqnarray}
where $i+1$ marks the $i+1$th grid cell. The total flux emitted by the star is given by $L_\star=4\pi R_\star^2 \sigma T_\star^4$. The stellar flux per surface area on a sphere with radius $r$ is thus
\begin{equation}
 F_\star (r) = \frac{L_\star}{4\pi r^2} = \sigma T_\star^4 \left(\frac{R_\star}{r}\right)^2 \ .
\end{equation}
With the front area of a grid cell given by
\begin{equation}
  A = r^2 \Delta \varphi (\cos \theta_1 - \cos \theta_2)
\end{equation}
we can compute the flux received by a single grid cell
\begin{equation}
  F_{\star} \cdot A = R_\star^2  \sigma T_\star^4 \Delta \varphi (\cos \theta_1 - \cos \theta_2) = F_{S \star} \ ,
\end{equation}
which then leads to the stellar heating $s$ of a grid cell
\begin{eqnarray}
  s &=& F_{S \star} e^{-\tau_i} \left(1 - e^{-\rho_i \kappa_i \Delta r} \right) \\ &=& R_\star^2  \sigma T_\star^4 \Delta \varphi (\cos \theta_1 - \cos \theta_2) e^{-\tau_i} \left(1 - e^{-\rho_i \kappa_i \Delta r} \right) \ .
\end{eqnarray}
As the energy equation (eq.~\ref{eq:energy}) is written for energy densities, we have to divide $s$ by the volume of the grid cell $V=r^2 \Delta r \Delta \varphi (\cos \theta_1 - \cos \theta_2)$ to get the same units and ultimately to get $S=s/V$ as it is used in the code:
\begin{equation}
\label{eqa:stellheat}
  S = F_\star  e^{-\tau_i} \frac{1-e^{-\rho \kappa_\star \Delta r}}{\Delta r}
\end{equation}
This result yields a dependence on resolution. However, this is only important for the optically thick regions ($\rho \kappa_\star \Delta r>1$); for optically thin regions we make the approximation $(1 - e^{-\rho \kappa_\star \Delta r})/(\Delta r)=\rho \kappa_\star$ . In the inner, optically thick regions of the disc the resolution influences the repartition of the stellar heating, but it does not influence the global net heating. Resolution studies indicate that the height of the puffed up inner edge varies by $\approx 3\%$ when lowering the resolution by $33\%$. As the disc is radially optically thick, the stellar flux is absorbed in the first cells; beyond that the simulations match perfectly for all tested resolutions.

\subsection{Innermost bumps and relation to opacity}
\label{ap:Hrdrop}

\begin{figure}
 \centering
 \resizebox{\hsize}{!}{\includegraphics[width=0.9\linwx]{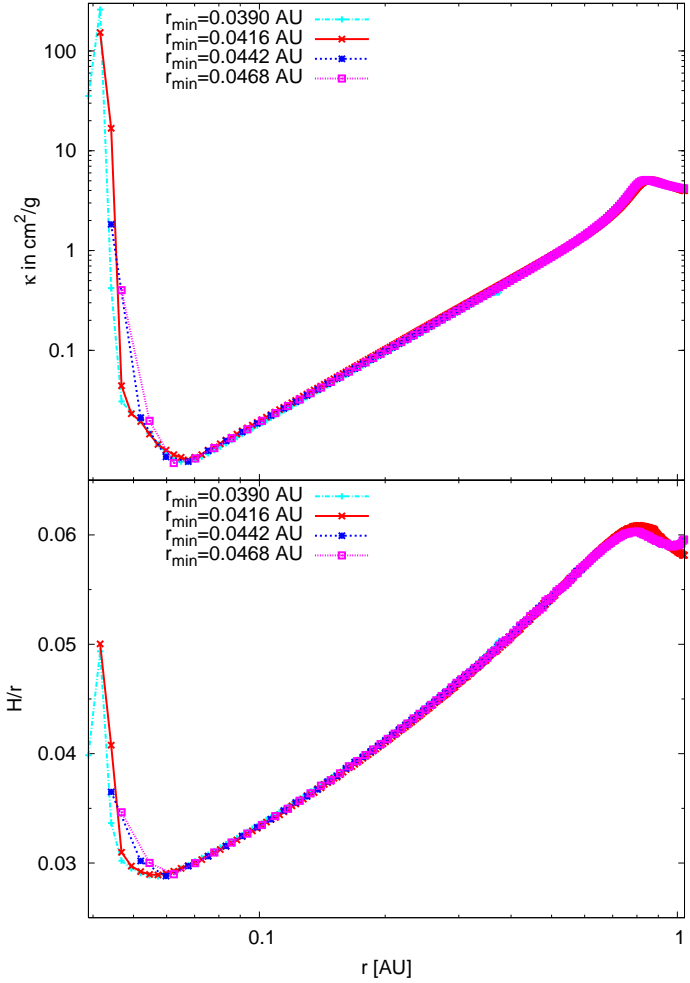}}
 \caption{Inner bump of the disc featuring different $r_{min}$ for the same disc configuration as in Fig.~\ref{fig:Hrinnernck}. The top panel shows $\kappa$, while $H/r$ is shown in the bottom panel. 
   \label{figA:Hrabsorb}
   }
\end{figure}

The innermost bump shown in Fig.~\ref{fig:Hrinnernck} and Fig.~\ref{fig:Hralphainnernck} can be mistaken as a puffed up inner rim caused by stellar irradiation, which would be in contradiction to what is stated in eq.\ref{eq:Hrinner}. However, the temperature in that region is a few $1000$K. At that temperature the opacity profile shows a kink (see Fig.~\ref{fig:opacity}), indicating that the disc structure will change. In Fig.~\ref{fig:Hrinnernck} the change of $H/r$ can be related to a change in opacity. As the opacity drops for slightly larger distances than $r_{min}$, so too does $H/r$. As the opacity determines the disc structure, we show in Fig.~\ref{figA:Hrabsorb} simulations of the inner disc with different $r_{min}$ for the same disc configuration as in Fig.~\ref{fig:Hrinnernck}.

For distances larger than $0.0416$AU, the opacity drops (top in Fig.~\ref{figA:Hrabsorb}) and therefore $H/r$ drops as well (bottom in Fig.~\ref{figA:Hrabsorb}). For distances closer to the star, the opacity drops (as already indicated by Fig.~\ref{fig:opacity}, because of the higher temperature) and $H/r$ decreases again. This clearly shows that the innermost bump is related to a transition in opacity, as the aspect ration of the disc drops for smaller and larger distances from $0.0416$AU.

As the opacity for even larger temperatures decreases continuously (due to the assumed power-law in the opacity for these temperatures) our simulations cover the innermost rim, indicating that no additional rim is inside.

\subsection{Matching of inner and outer disc}
\label{ap:match}

\begin{figure}
 \centering
 \resizebox{\hsize}{!}{\includegraphics[width=0.9\linwx]{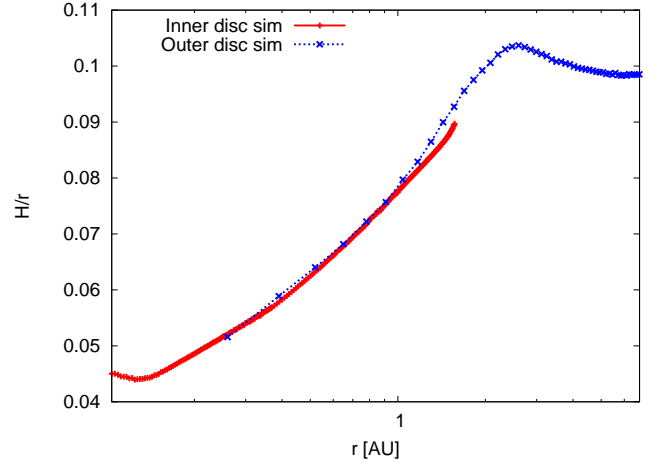}}
 \caption{Matching of the inner disc simulations with the outer disc simulations for an $\alpha=0.008$ disc with $\Sigma_0=3000$g/cm$^2$.
   \label{figA:Hrmatch}
   }
\end{figure}

The models of the inner and outer disc can be attached to each other. In order to do so, the regions of the simulations have to overlap. In Fig.~\ref{figA:Hrmatch} we present the overlapping region of the $\alpha=0.008$ simulation with $\Sigma_0=3000$g/cm$^2$, which are actually the simulations presented in Fig.~\ref{fig:Hralphainnernck} and Fig.~\ref{fig:Hralphaouternck}.

Clearly $H/r$ of the simulations of inner and outer disc match very well. We are therefore confident that the approximation of $\theta_{abs}$, as presented in eq.~\ref{eq:thetaabs}, for the stellar heating is sufficient. If the matching was done in a region of the disc that is dominated by stellar irradiation, the matching might fall into a shadowed region of the disc that would then not be captured by the outer disc leading to a mismatch of the two simulations. However, if, as in our simulations the matching occurs in a viscously dominated region of the disc, the outer discs inner edge will adapt to the outer edge of the inner edge due to the viscous heating.

\section{Energy equation}
\label{ap:energy}

For simplicity the derivation of solving the energy equation (eq.~\ref{eq:energy}) is shown in Cartesian coordinates in the implemented 3D algorithm, although we used 2D simulations in the $r$-$\theta$ plane in this work. The radiation energy equation is then given by
\begin{eqnarray}
  \label{apBeq:energy}
  \frac{E_R^{n+1} -E_R^n}{\Delta t} &= \frac{1}{\Delta x} \left( \bar{D}_{i+1,j,k}^x \frac{E_{R,i+1,j,k}^{n+1} - E_{R,i,j,k}^{n+1}}{\Delta x}
 -\bar{D}_{i,j,k}^x \frac{E_{R,i,j,k}^{n+1} - E_{R,i-1,j,k}^{n+1}}{\Delta x} \right) \nonumber \\
 &+ \frac{1}{\Delta y} \left( \bar{D}_{i,j+1,k}^y \frac{E_{R,i,j+1,k}^{n+1} - E_{R,i,j,k}^{n+1}}{\Delta y}
 -\bar{D}_{i,j,k}^y \frac{E_{R,i,j,k}^{n+1} - E_{R,i,j-1,k}^{n+1}}{\Delta y} \right) \nonumber \\
 &+ \frac{1}{\Delta z} \left( \bar{D}_{i,j,k+1}^z \frac{E_{R,i,j,k+1}^{n+1} - E_{R,i,j,k}^{n+1}}{\Delta z}
 -\bar{D}_{i,j,k}^z \frac{E_{R,i,j,k}^{n+1} - E_{R,i,j,k-1}^{n+1}}{\Delta z} \right) \nonumber \\
 &+ \rho \kappa_P [ 4 \sigma (T^{n+1})^4 - c E_R^{n+1}] \ ,
\end{eqnarray}
where the upper indices ($n$ and $n+1$) indicate the time step and 
\begin{eqnarray}
 \bar{D}_{i,j,k}^x &= \frac{1}{2} \left( D_{i,j,k} + D_{i-1,j,k} \right) \ , \nonumber \\
 \bar{D}_{i,j,k}^y &= \frac{1}{2} \left( D_{i,j,k} + D_{i,j-1,k} \right) \ , \nonumber \\
 \bar{D}_{i,j,k}^z &= \frac{1}{2} \left( D_{i,j,k} + D_{i,j,k-1} \right) \ , \nonumber
\end{eqnarray}
are the averaged diffusion coefficients, which are given by
\begin{equation}
 D_{i,j,k} = \frac{\lambda c}{\rho_{i,j,k} \kappa_{R,i,j,k}} \ .
\end{equation}
The term $(T^{n+1})^4$ is non linear and makes the scheme difficult to invert, yet it is much easier to solve a linear system implicitly. Assuming that the changes in temperature are small in each time step \citep{2011A&A...529A..35C}, we can write
\begin{equation}
\label{apBeq:Temp}
  (T^{n+1})^4 = 4 (T^n)^3 T^{n+1} - 3 (T^n)^4 \ .
\end{equation}
Now, we need to have $T^{n+1}$ as a function of $T^n$, $E_R^n$ and $E_R^{n+1}$ in order to solve equation.~\ref{apBeq:energy}. We use now the energy density equation, where we omit advection and compressional heating,
\begin{equation}
 \doverd{\, \epsilon}{t} =  - \rho \kappa_P (T,P) [B(T) - c E_R] + Q^+ + S \ .
\end{equation}
With $\epsilon=\rho c_v T$ we get
\begin{equation}
 \frac{T^{n+1} - T^n}{\Delta t} = - \frac{\kappa_P}{c_V} (4 \sigma (T^{n+1})^4 - c E_R^{n+1}) + \frac{S}{\rho c_V}  + \frac{Q^+}{\rho c_V} \ . 
\end{equation}
With our approximation for the temperature (eq.~\ref{apBeq:Temp}) we find
\begin{equation}
\label{apBeq:eta}
 T^{n+1} = \eta_1 + \eta_2 E_R^{n+1} \ ,
\end{equation}
where
\begin{eqnarray}
  \eta_1 &=& \frac{T^n + 12 \Delta t \frac{\kappa_P}{c_V} \sigma (T^n)^4 + \frac{\Delta t S}{\rho c_V} + \frac{\Delta t Q^+}{\rho c_V}}{1 + 16 \Delta t \frac{\kappa_P}{c_V} \sigma (T^n)^3 } \quad \mathrm{and} \\
 \eta_2 &=& \frac{\Delta t \frac{\kappa_P}{c_V} c}{1 + 16 \Delta t \frac{\kappa_P}{c_V} \sigma (T^n)^3 } \ .
\end{eqnarray}
We can now plug this all into the radiation energy equation and solve for $E_R^{n+1}$. We arrive at a matrix equation:
\begin{eqnarray}
  \beta_{1,i,j,k} E_{R,i+1,j,k} + \beta_{2,i,j,k} E_{R,i-1,j,k} + \beta_{3,i,j,k} E_{R,i,j+1,k} &+& \beta_{4,i,j,k} E_{R,i,j-1,k} \nonumber \\
  + \beta_{5,i,j,k} E_{R,i,j,k+1} + \beta_{6,i,j,k} E_{R,i,j,k-1} + \Gamma_{i,j,k} E_{R,i,j,k} &=& R_{i,j,k} \nonumber \ ,
\end{eqnarray}
where the superscript $n+1$ has been omitted on the left hand side. The matrix elements are given by:
\begin{eqnarray}
  \beta_{1,i,j,k} &=& - \frac{\Delta t}{\Delta x^2} \bar{D}_{i+1,j,k}^x \nonumber \\
 \beta_{2,i,j,k} &=& - \frac{\Delta t}{\Delta x^2} \bar{D}_{i-1,j,k}^x \nonumber \\
 \beta_{3,i,j,k} &=& - \frac{\Delta t}{\Delta y^2} \bar{D}_{i,j+1,k}^y \nonumber \\
 \beta_{4,i,j,k} &=& - \frac{\Delta t}{\Delta y^2} \bar{D}_{i,j-1,k}^y \nonumber \\
 \beta_{5,i,j,k} &=& - \frac{\Delta t}{\Delta z^2} \bar{D}_{i,j,k+1}^z \nonumber \\
 \beta_{6,i,j,k} &=& - \frac{\Delta t}{\Delta z^2} \bar{D}_{i,j,k-1}^z \nonumber \\
 \beta_{1-6} &=& -(\beta_1 + \beta_2 + \beta_3 + \beta_4 + \beta_5 + \beta_6) \nonumber \\
 \Gamma_{i,j,k}  &=& (1 + \Delta t \rho \kappa_{P} c - 16 \Delta t \rho \kappa_{P} \sigma (T^n)^3 \eta_2) + \beta_{1-6} \nonumber \\
 R_{i,j,k} &=& 16 \Delta t \rho \kappa_{P} \sigma (T^n)^3 \eta_1 - 12 \Delta t \rho \kappa_{P} \sigma (T^n)^4 + E_R^n \nonumber
\end{eqnarray}
Written in matrix notation we have
\begin{equation}
 M \mathbf{E_R}^{n+1} = \mathbf{R} \ ,
\end{equation}
which is a matrix that is very similar in nature to the one for the one-temperature energy equation and can be solved with the same matrix solver as in \citet{2009A&A...506..971K}. After the matrix iteration, the new temperature of the system can be calculated with eq.~\ref{apBeq:eta}, followed by the computation of $\epsilon = \rho c_v T$.

\bibliographystyle{aa}
\bibliography{Stellar}
\end{document}